\documentclass[amssymb,amsmath,twocolumn,prb]{revtex4}

\usepackage{color,graphicx,bm}
\usepackage{upgreek}
\usepackage{overpic}
\usepackage{subfigure}

\newcommand{\etal}{{\it et al}.}
\newcommand{\beq}{\begin{equation}}
\newcommand{\eeq}{\end{equation}}
\newcommand{\ua}{\uparrow}
\newcommand{\da}{\downarrow}

\begin{document}

\title{Flux-Periodicity Crossover from $\bm{hc/e}$ in Normal Metallic to $\bm{hc/2e}$ in Superconducting Loops} 

\author{Florian Loder$^{\,1,2}$, Arno P. Kampf$^{\,2}$, and Thilo Kopp$^{\,1}$\vspace{0,3cm}}

\affiliation{Center for Electronic Correlations and Magnetism, $^1$Experimental Physics VI, $^2$Theoretical Physics III\\ 
Institute of Physics, University of Augsburg, 86135 Augsburg, Germany}

\date{\today}

\begin{abstract}
The periodic response of a metallic or a superconducting ring to an external magnetic flux is one of the most evident manifestations of quantum mechanics. It is generally understood that the oscillation period $hc/2e$ in the superconducting state is half the period $hc/e$ in the metallic state, because the supercurrent is carried by Cooper pairs with a charge $2e$. On the basis of the Bardeen-Cooper-Schrieffer theory we discuss, in which cases this simple interpretation is valid and when a more careful analysis is needed. In fact, the knowledge of the oscillation period of the current in the ring provides information on the electron interactions. In particular, we analyze the crossover from the $hc/e$ periodic normal current to the $hc/2e$ periodic supercurrent upon turning on a pairing interaction in a metal ring. Further, we elaborate on the periodicity crossover when cooling a metallic loop through the superconducting transition temperature $T_{\rm c}$.\\[3mm]
\end{abstract}

\maketitle

\section{Introduction}

One of the most important properties of superconductors is their perfectly diamagnetic response to an external magnetic field, the Meissner-Ochsenfeld effect. It is a pure quantum effect and therefore reveals the existence of a macroscopic quantum state with a pair condensate. A special manifestation of the diamagnetic response is observed for superconducting rings threaded by a magnetic flux: flux quantization and a periodic current response.

Persistent currents and periodic flux dependence are also known in normal metal rings and best known in form of the Aharonov-Bohm effect predicted theoretically in 1959~\cite{AB}. Since the wavefunction of an electron moving on a ring must be single valued, the phase of the wave function acquired upon moving once around the ring is a integer multiple of $2\pi$. A magnetic flux threading the ring generates an additional phase difference $2\pi\varphi=e/(\hslash c)\oint_{C}{\rm d}{\bf r}\cdot{\bf A}({\bf r})=[2\pi e/(hc)]\,\varphi$, where $C$ is a closed path around the ring and ${\bf A}({\bf r})$ the vector potential generating the magnetic flux $\varphi$ threading the ring. Here, $e$ is the electron charge, $c$ the velocity of light, and $h$ is Planck's constant. Thus, the electron wave function is identical whenever $\varphi$ has an integer value and therefore the system is periodic in the magnetic flux $\varphi$ with a periodicity of
\begin{align}
\Phi_0=\frac{hc}{e},
\end{align}
the flux quantum in a normal metal ring. In particular, the persistent current $J(\varphi)$ induced by the magnetic flux is zero whenever $\Phi=\varphi/\Phi_0$ is an integer.

The periodic response of a superconducting ring to a magnetic flux is of similar origin as in a normal metal ring, though the phase winding of the condensate wavefunction has to be reconsidered. On account of the macroscopic phase coherence of the condensate, flux oscillations must be more stable in superconductors, and London predicted their existence in superconducting loops already ten years before the work of Aharonov and Bohm~\cite{London}. London expected that the magnetic flux threading a loop is quantized in multiples of $\Phi_0$ because the interior of an ideal superconductor was known to be current free. 
Although London the pairing theory of superconductivity was not known yet, he anticipated the existence of electron pairs carrying the supercurrent and speculated that the flux quantum in a superconductor might be $\Phi_0/2$.
This point of view became generally accepted with the publication of the \textquoteleft Theory of Superconductivity' by Bardeen, Cooper, and Schrieffer (BCS) in 1957~\cite{bcs}. Direct measurements of magnetic flux quanta $\Phi_0/2$ trapped in superconducting rings followed in 1961 by Doll and N\"abauer~\cite{Doll} and by Deaver and Fairbank~\cite{Deaver}, corroborated later by the detection of flux lines of $\Phi_0/2$ in the mixed state of type II superconductors~\cite{Abrikosov,Essmann}. 

It is tempting to explain the $\Phi_0/2$ flux periodicity of superconducting loops simply by the charge $2e$ of Cooper pairs carrying the supercurrent, but pairing of electrons alone is not sufficient for the $\Phi_0/2$ periodicity. The Cooper-pair wavefunction extends over the whole loop, as does the single-electron wavefunction, and it is not obvious whether the electrons forming the Cooper pair are tightly bound or circulate around the ring separately. A microscopic model on the basis of the BCS theory is therefore indispensable for the description of the flux periodicity of a superconducting ring. In this chapter, we analyze this problem in detail and focus on a previously neglected aspect: how do the $\Phi_0$ periodic flux oscillations in a normal metal ring transform into the $\Phi_0/2$ periodic oscillations in a superconducting ring?

\begin{figure*}[t]	
\centering
\vspace{3mm}
\begin{overpic}
[width=0.7\columnwidth]{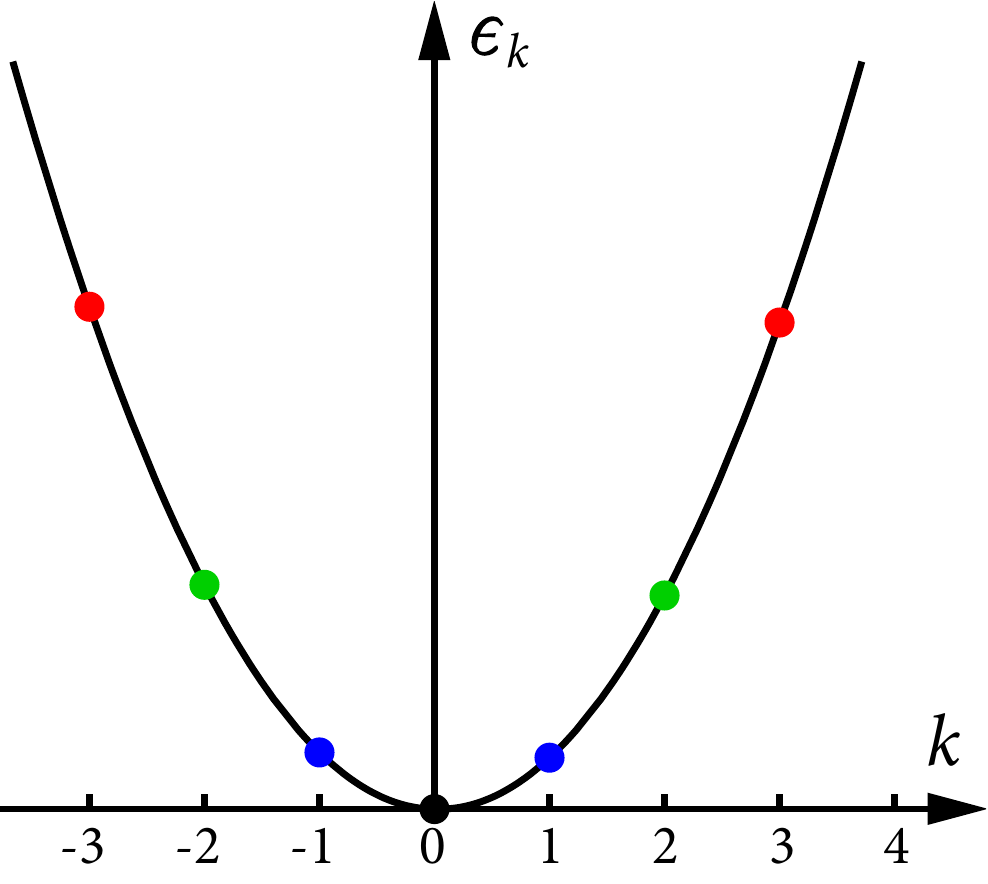}
\put(-10,80){(a)}
\end{overpic}
\hspace{18mm}
\begin{overpic}
[width=0.7\columnwidth]{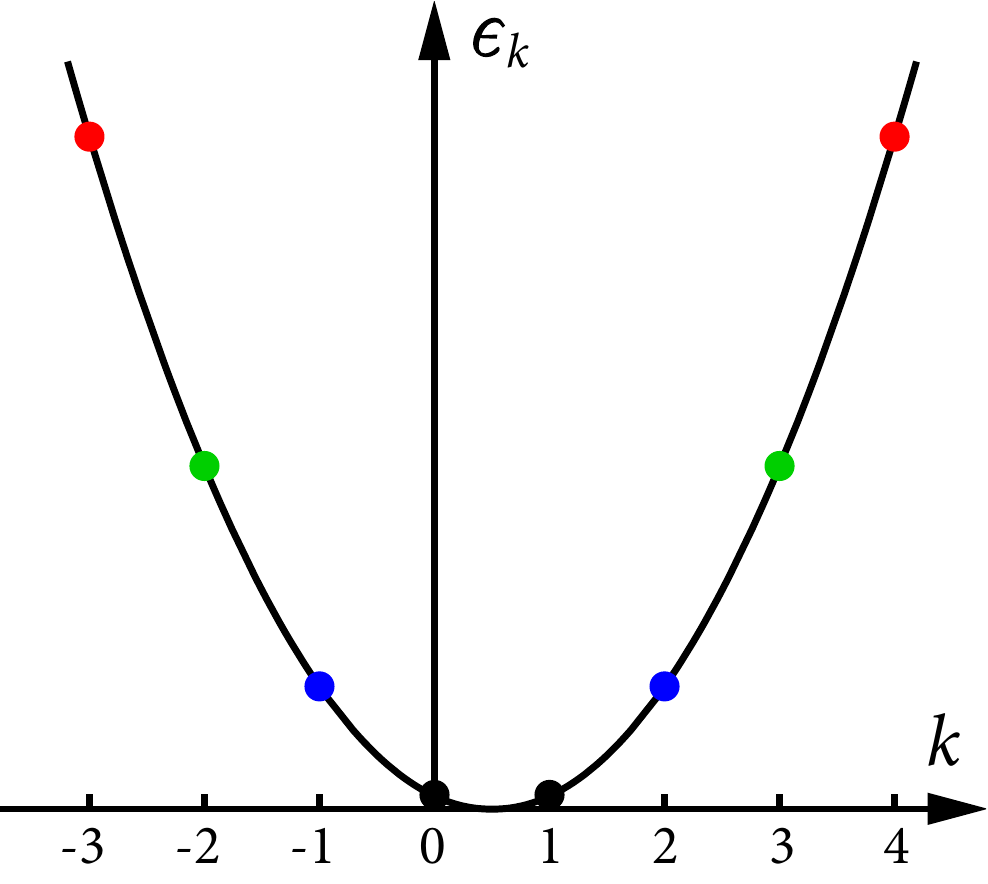}
\put(-6,80){(b)}
\end{overpic}\\[0mm]
\caption{Scheme of the pairing of angular-momentum eigenstates in a one dimensional metal loop for (a) $\Phi=0$ and (b) $\Phi=\Phi_0/2$, as used by Schrieffer in~\cite{schrieffer} to illustrate the origin of the $\Phi_0/2$ periodicity in superconductors. Paired are always electrons with equal energies, leading to center-of-mass angular momenta $q=0$ in (a) and $q=1$ in (b).}
\label{fig:Sub_00}
\end{figure*}

A theoretical description of the origin of the half-integer flux quanta was first found independently in 1961 by Byers and Yang~\cite{Byers}, by Onsager~\cite{onsager:61}, and by Brenig~\cite{brenig:61} on the basis of BCS theory. They realized that there are two distinct classes of superconducting wavefunctions that are not related by a gauge transformation. An intuitive picture illustrating these two types can be found in Schrieffer's book on superconductivity~\cite{schrieffer}, using the energy spectrum of a one-dimensional metal ring. The first class of superconducting wavefunctions, which London had in mind in his considerations about flux quantization, is related to pairing of electrons with angular momenta $\hslash k$ and $-\hslash k$, which have equal energies in a metal loop without magnetic flux, as schematically shown in figure~\ref{fig:Sub_00}~(a). The Cooper pairs in this state have a center-of-mass angular momentum (pair momentum) $\hslash q=0$. The pairing wavefunctions of the superconducting state for all flux values $\Phi$, which are integer multiples of $\Phi_0$ and correspond to even pair momenta $\hslash q=2\hslash\Phi/\Phi_0$, are related to the wavefunction for $\Phi=0$ by a gauge transformation. For a flux value $\Phi_0/2$, pairing occurs between the electron states with angular momenta $\hslash k$ and $\hslash(-k+1)$, which have equal energies in this case [figure~\ref{fig:Sub_00}~(b)]. This leads to pairs with momentum $\hslash q=\hslash$. The corresponding pairing wavefunction is again related by a gauge transformation to those for flux values $\Phi$ which are half-integer multiples of $\Phi_0$ and correspond to the odd pair momenta $\hslash q=2\hslash\Phi/\Phi_0$.

For the system to be $\Phi_0/2$ periodic, it is required that the free energies of the two types of pairing states are equal. Byers and Yang, Onsager as well as Brenig showed that this is in fact the case in the thermodynamic limit. The free energy consists then of a series of parabolae with minima at integer multiples of $\Phi_0$ (corresponding to even pair momenta) and half integer multiples of $\Phi_0$ (corresponding to odd pair momenta). If the arm of the ring is wider than the penetration depth $\lambda$, the flux is quantized and the groundstate is given by the minimum closest to the value of the external flux. However, in microscopic finite systems this degeneracy of the even and odd $q$ minima is lifted, although their position is fixed by gauge invariance to multiples of $\Phi_0/2$. The restoration of the $\Phi_0/2$ periodicity in the limit of large rings was studied only much later~\cite{zhu:94,khavkine:04,loder:08.2,vakaryuk:08}. We study the revival of the $\Phi_0/2$ periodicity in sections~\ref{sec:norm} and \ref{sec:em} for a one-dimensional ring at zero temperature and investigate the effects of many channels and finite temperatures in section~\ref{sec:self}.

From the flux periodicity of the free energy, the same flux periodicity can be derived for all other thermodynamic quantities~\cite{Tinkham}. A clear and unambiguous observation of flux oscillations is possible in the flux dependence of the critical temperature $T_{\rm c}$ of small superconducting cylinders. Such experiments have been performed first by Little and Parks in 1962~\cite{Little,Parks,little:64}. They measured the resistance $R$ of the cylinder at a fixed temperature $T$ within the finite width of the superconducting transition and deduced the oscillation period of $T_{\rm c}$ from the variation of $R$. These experiments confirmed the $\Phi_0/2$ periodicity in conventional superconductors very accurately.
At this stage the question of the flux periodicity in superconductors seemed to be settled and understood. The interest then shifted to the amplitude of the supercurrent and also the normal persistent current and their dependence on the ring size, the temperature, and disorder~\cite{Landauer, landauer:85, cheung:88, oppen:91,eckern}. However, the influence of finite system sizes on the flux periodicity remained unaddressed. Earlier, certain experiments had already indicated some unexpected complications.
E.g., Little and Parks pointed out in reference~\cite{little:64} that in tantalum cylinders they could not detect any flux oscillations in $R$ at all. Even more peculiar were the oscillations observed in an indium cylinder where signs of an additional $\Phi_0/8$ periodicity were clearly visible~\cite{little:64}. This was surprising because indium is a perfectly conventional superconductor otherwise. These results remained unexplained and drew attention only years later, when flux oscillations of unconventional superconductors were studied.

\begin{figure*}[t]	
\centering
\vspace{3mm}
\begin{overpic}
[width=0.8\columnwidth]{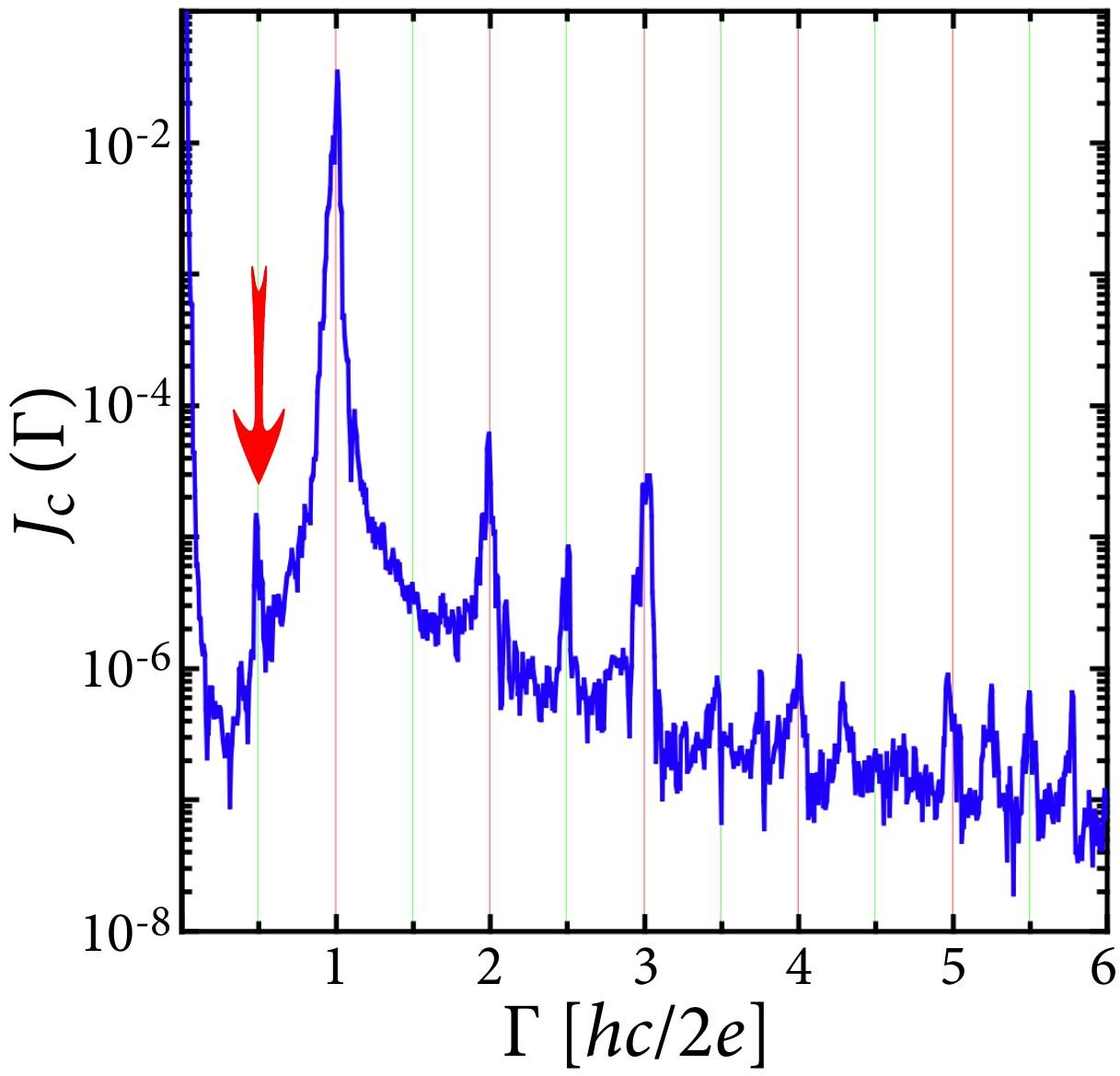}
\put(-6,92){(a)}
\end{overpic}
\hspace{15mm}
\begin{overpic}
[width=0.8\columnwidth]{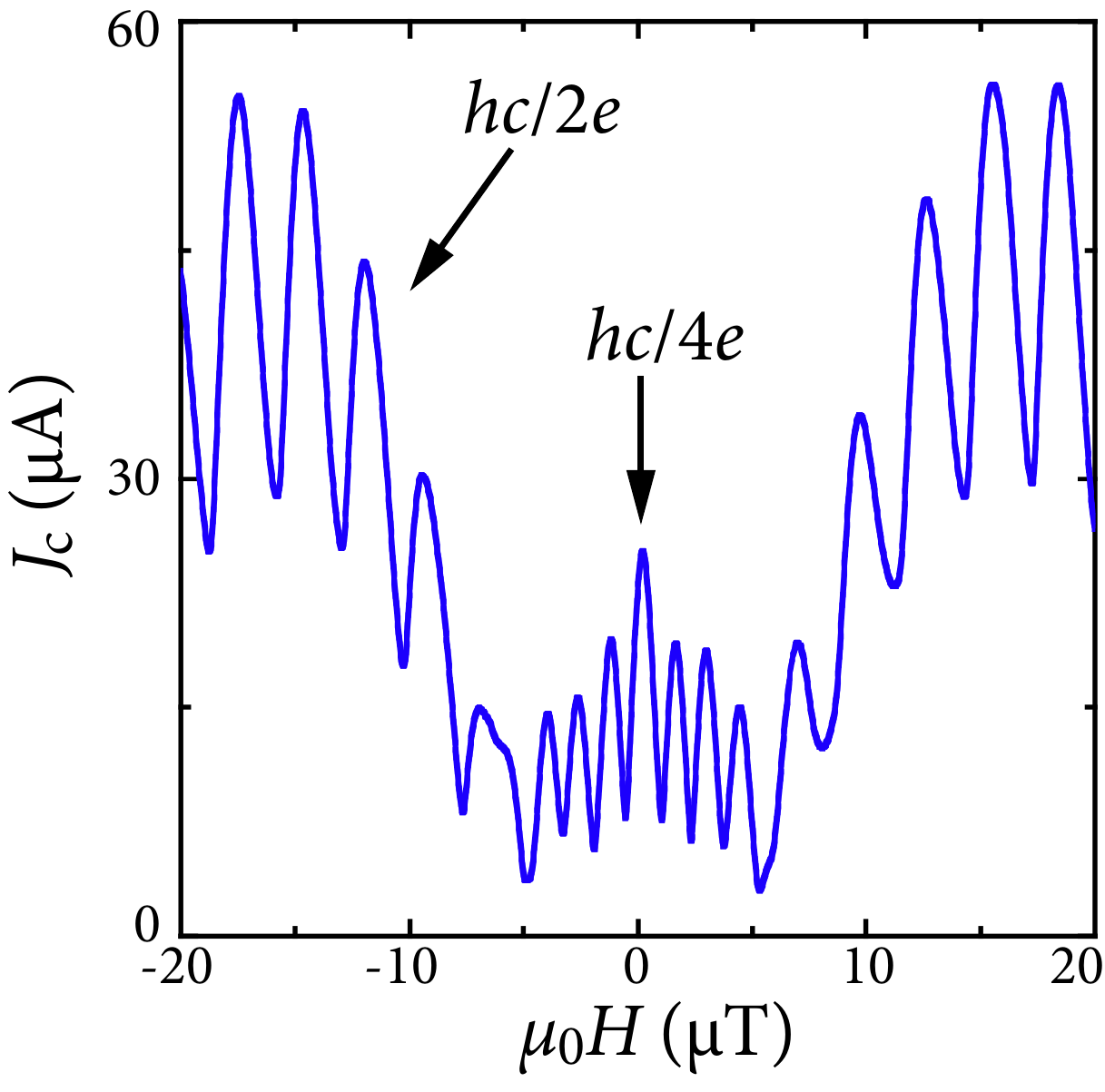}
\put(-6,92){(b)}
\end{overpic}\\[0mm]
\caption{(a) Fourier transform $J_{\rm c}(\Gamma)$ of the critical current $J_{\rm c}(H)$ measured by Schneider \etal\ on a 24$^\circ$ grain boundary SQUID at $T=$ 77$\,$K as a function of the applied magnetic field where $\Phi_0/2=6.7\,\upmu$T~\cite{schneider}. (b) Critical current $J_{\rm c}(H)$ over a 24$^\circ$ grain boundary SQUID at $T=4.2\,$K, where  $\Phi_0/2=2.7\,\upmu$T. Clearly visible is the abrupt change of periodicity at $\mu_0H\approx\pm5\,\upmu$T~\cite{schneider:04}.}
\label{fig:Sub_0}
\end{figure*}

In the meantime a new type of flux sensitive systems was advanced: superconducting quantum-interference devices (SQUIDs). The measurement of flux oscillations in SQUIDs is similar to the Little-Parks experiment. Here the flux dependence of the critical current $J_{\rm c}$ through a superconducting loop including one or two Josephson junctions is measured. This has the advantage that flux oscillations can be observed at any temperature $T<T_{\rm c}$, and they are most clearly visible in the critical current $J_{\rm c}$. 
SQUIDs fabricated from conventional superconductors have been used in experiments and applications for five decades, and they proved to oscillate perfectly with the expected flux period $\Phi_0/2$. It was therefore a surprise that flux oscillations with different periodicities were found  in 2003 by Lindstr\"om \etal~\cite{lind:03} and Schneider \etal~\cite{schneider:04,schneider} in SQUIDs fabricated from films of the high-$T_{\rm c}$ superconductor YBa$_2$Cu$_3$O$_y$ (YBCO) where the Josephson junctions arise from grain boundaries. Flux trapping experiments in loops showed that flux quantization in the cuprate class of high-$T_{\rm c}$ superconductors occurs in units of $\Phi_0/2$~\cite{Gough}, identically to what has been observed with conventional superconductors. In addition, Schneider \etal\ observed a variety of oscillation periods, depending on the geometry of the SQUID loop, the grain-boundary angle, the temperature, and the magnetic-field range of the SQUID.

Two distinct patterns of unconventional oscillations in YBCO SQUIDs have to be discerned. The first kind consists of oscillations which have a basic period of $\Phi_0/2$, overlaid by other periodicities, such that the Fourier transform $J_{\rm c}(\Gamma)$ of $J_{\rm c}(\Phi)$ contains peaks appear which do not correspond to the period $\Phi_0/2$~\cite{schneider}. An example for such a measurement are shown in figure~\ref{fig:Sub_0}~(a). The peaks at integer values of $\Gamma$ correspond to higher harmonics of $\Phi_0/2$, and their appearance is natural. However, there are clear peaks at $\Gamma=1/2$ (red arrow) and $\Gamma=5/2$, which correspond to $\Phi_0$ periodicity and higher harmonics thereof. The origin of the $\Phi_0$ periodicity in those experiments is so far not conclusively explained. There was, however, extensive research on the flux periodicity of unconventional (mostly $d$-wave) superconductors, which revealed that the periodicity of the normal state persists in the superconducting state if the energy gap symmetry allows for nodal states~\cite{loder:08,barash:07,tesanovic:08,loder:09,zha:09}. This effect derives directly from the analysis in this book chapter and is discussed in detail in reference~\cite{loder:09}.

The second kind of unconventional oscillations is more intriguing. In several different YBCO SQUIDs, the periodicity of sinusoidal oscillations changes abruptly with increasing magnetic flux. In the measurement shown in figure~\ref{fig:Sub_0}~(b), the period is $\Phi_0/4$ for small flux, and changes to $\Phi_0/2$ at a critical flux. As a possible explanation for the appearance of $\Phi_0/4$ periodicity, an unusually pronounced second harmonic in the critical current $J_{\rm c}$ of transparent Josephson junctions was proposed or, more fundamentally, an effect of interactions between Cooper pairs, leading to the formation of electron quartets~\cite{schneider:04}. The observation of similar abrupt changes to other fractional periodicities like $\Phi_0/6$ and $\Phi_0/8$ render this finding even more striking since it could indicate a transition into a new, non-BCS type of superconductivity. This concept, which we sketch briefly in the Conclusions, is a complex and promising topic for future research on unconventional superconductors.

\section{The Periodicity Crossover}\label{sec:cross}

In this section we introduce the periodicity crossover and consider first the simplest model containing the relevant physics: a one dimensional ring consisting of $N$ lattice sites and a lattice constant $a$ (figure~\ref{fig:Sub_1}). The ring is threaded by a magnetic flux $\Phi$ which does not touch the ring itself. We use a tight-binding description with nearest-neighbor hopping parameter $t$, which sets the energy scale of the system. We start from the flux periodicity of the normal metal state of the ring, which varies for different numbers of electrons in the ring. On this basis we introduce a superconducting pairing interaction and investigate the flux periodicity of the groundstate upon increasing the interaction strength. For a ring with a finite width (an annulus) we investigate the flux dependence of the self-consistently calculated superconducting order parameter and study the temperature driven periodicity crossover when cooling the ring through the transition temperature $T_{\rm c}$.

\subsection{Normal state}\label{sec:norm}

\begin{figure}[t]	
\centering
\includegraphics[width=0.6\columnwidth]{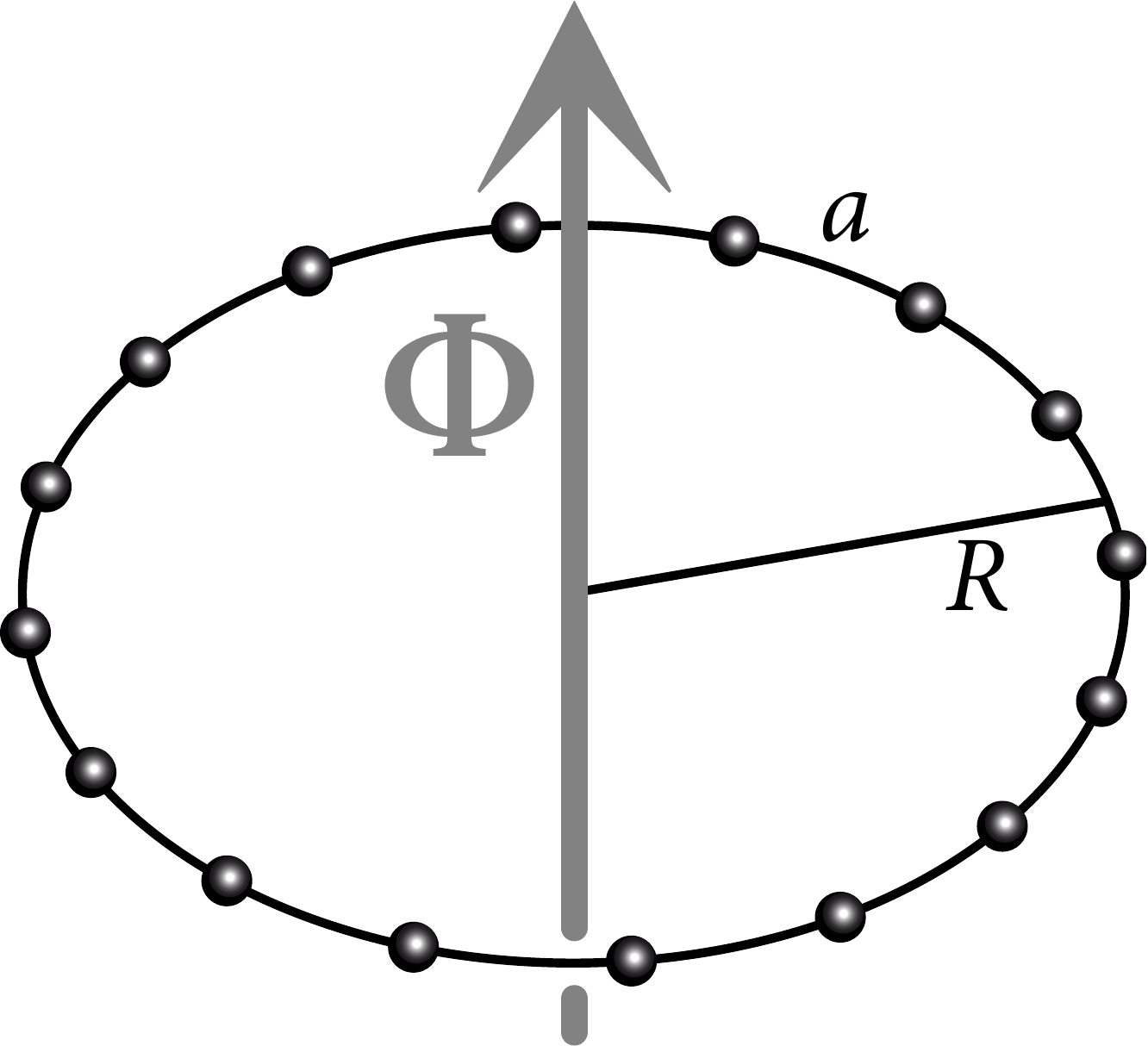}
\vspace{3mm}
\caption{The simplest description of the many-particle state in a flux threaded loop, we use a tight-binding model on a discrete, one-dimensional ring with $N$ lattice sites, lattice constant $a$ and radius $R=Na/2\pi$. The magnetic flux $\Phi$ is confined to the interior of the ring and does not touch the ring itself.}
\label{fig:Sub_1}
\end{figure}

The tight-binding Hamiltonian for an electronic system including a magnetic field is straightforwardly formulated using the annihilation and creation operators $c_{is}$ and $c^\dag_{is}$ for an electron with spin $s$ on the lattice site $i$:
\beq
{\cal H}_0=-t\sum_{\langle i,j\rangle,s}e^{i\varphi_{ij}}c_{is}^\dag c_{js}-\mu\sum_{i,s}c_{is}^\dag c_{is}.
\label{bcs01}
\eeq
Here $\langle i,j\rangle$ denotes all nearest-neighbor pairs $i$ and $j$, $s=\uparrow,\downarrow$. The magnetic field ${\bf B}=\bm\nabla\times{\bf A}$ enters into the Hamiltonian~(\ref{bcs01}) through the Peierls phase factor $
\varphi_{ij}=(e/hc)\int_i^jd{\bf l}\cdot{\bf A}$. The chemical potential $\mu$ controls the number of electrons in the ring. The flux periodicity is easiest to discuss for a particle-hole symmetric situation with $\mu=0$, for which the Fermi energy is $E_{\rm F}=0$. We will later address the changes introduced through an arbitrary $\mu$.

We assume that the $N$ lattice sites are equally spaced along a ring with circumference $2\pi R=Na$  (figure~\ref{fig:Sub_1}). It follows that the Peierls phase factor for a magnetic field focused through the center of the ring simplifies to $\varphi_{ij}=2\pi \varphi/N$, where $\varphi=\Phi/\Phi_0$ is the dimensionless magnetic flux. The Hamiltonian~(\ref{bcs01}) is then written in momentum space as:
\beq
{\cal H}_0=\sum_{k,s}\epsilon_k(\varphi)c_{ks}^\dag c_{ks}
\label{bcs04}
\eeq
where $c^\dag_{ks}$ creates an electron with angular momentum $\hslash k$. The energy dispersion is
\beq
\epsilon_k(\varphi)=-2t\cos\left(\frac{k-\varphi}{R/a}\right)-\mu.
\label{bcs05}
\eeq

\begin{figure}[t]
\centering
\vspace{3mm}
\begin{overpic}
[width=0.95\columnwidth]{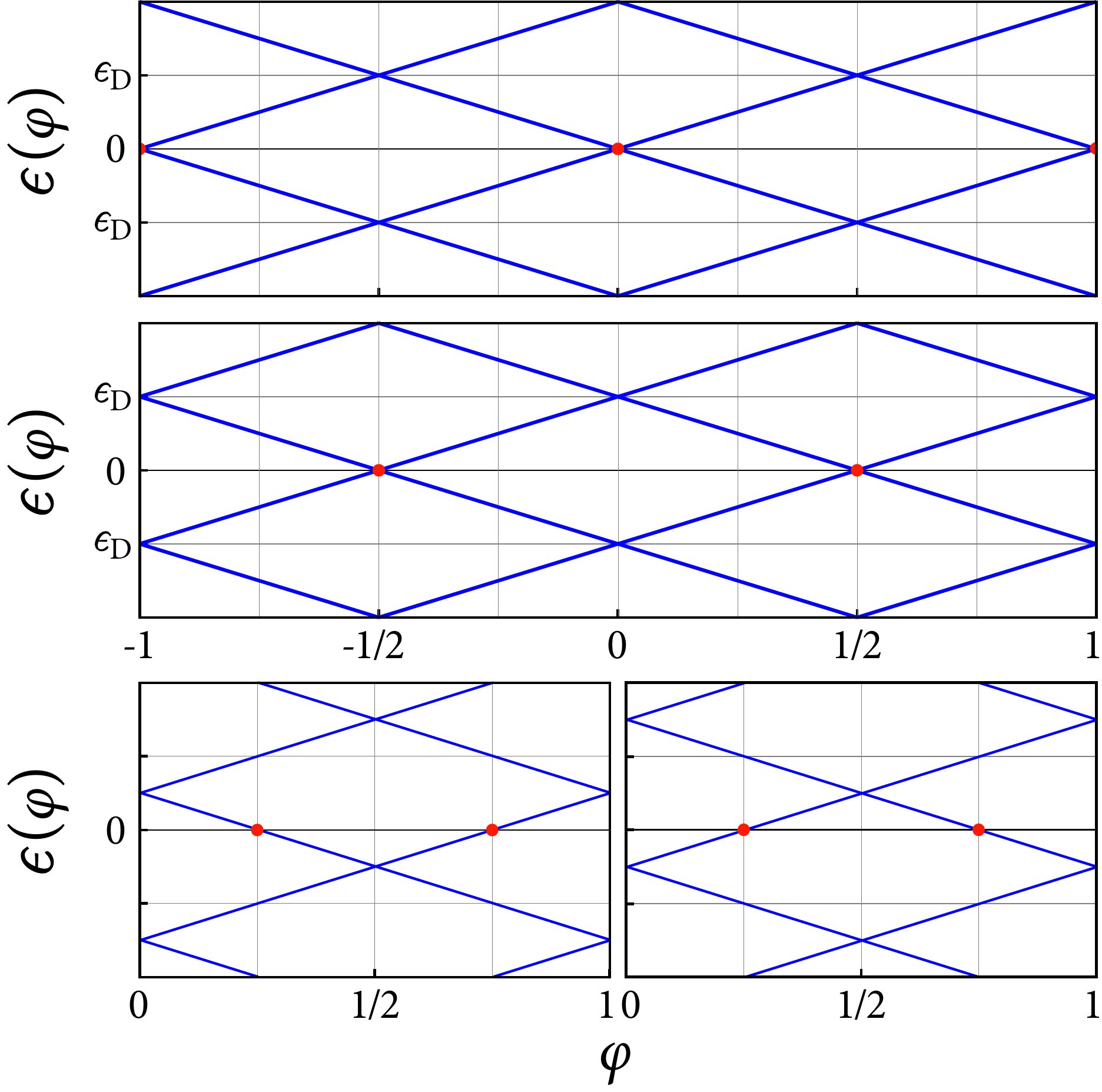}
\put(-3,95){\small{(a)}}
\put(-3,66.3){\small{(b)}}
\put(-3,33.5){\small{(c)}}
\end{overpic}
\caption{Energy spectrum of a discrete one-dimensional ring with $N$ lattice sites, $\mu=0$ and (a) $N/4$ is an integer, (b) $N/4$ a half integer, and (c) $N$ an odd number. In (a) and (b), levels cross $E_{\rm F}=0$ at integer (a) or half-integer (b) values of $\varphi$. For odd $N$, two different spectra are possible [$N=4n+1$ (left) and $N=4n-1$ (right)], and for both, two levels cross $E_{\rm F}$ within one flux period (red points). $\epsilon_{\rm D}$ denotes the maximum value of the Doppler shift.}
\label{Figs0.10}
\end{figure}

\begin{figure*}[t]	
\centering
\vspace{3mm}
\begin{overpic}
[width=0.95\columnwidth]{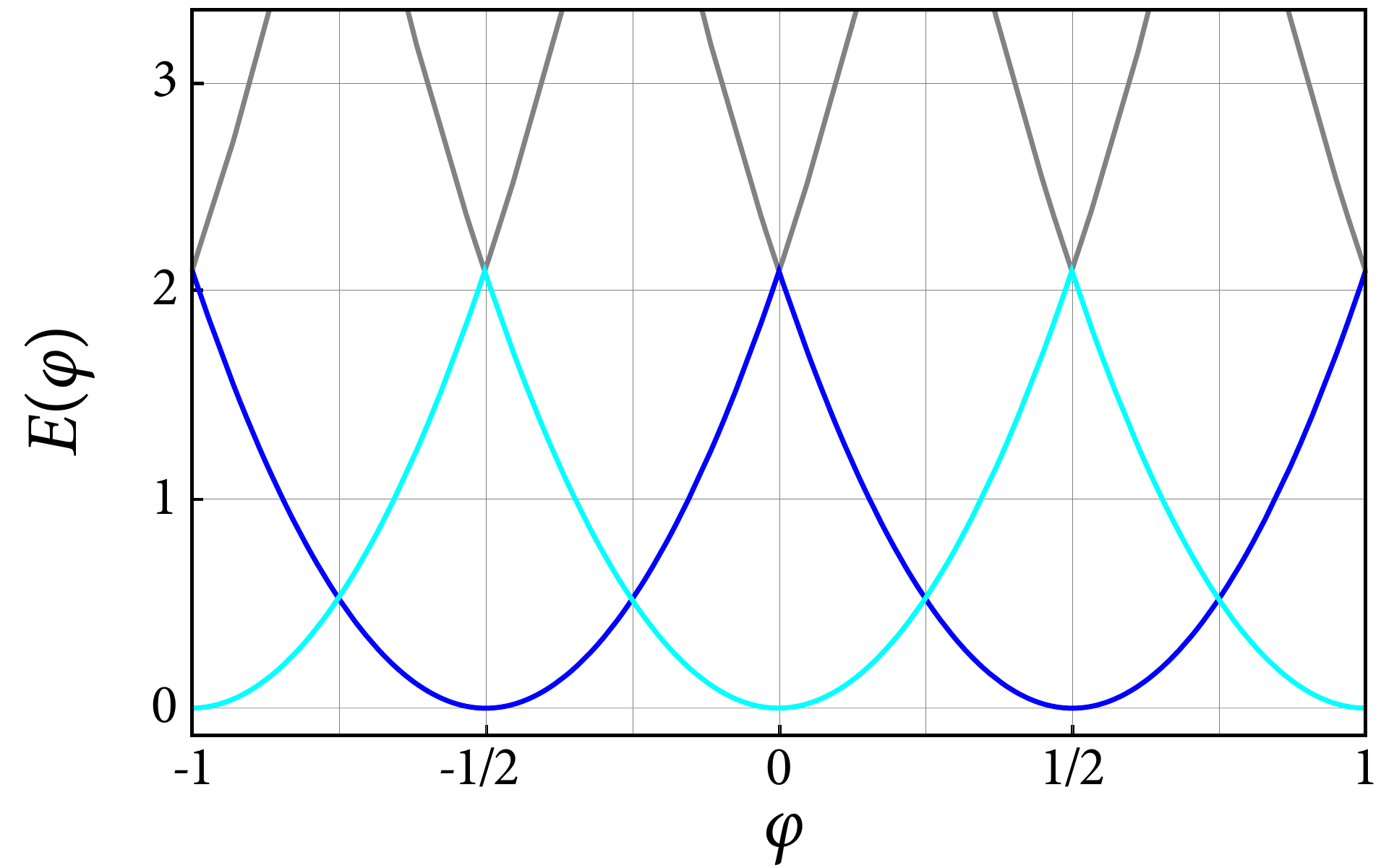}
\put(-3,59){(a)}
\end{overpic}
\hspace{6mm}
\begin{overpic}
[width=0.95\columnwidth]{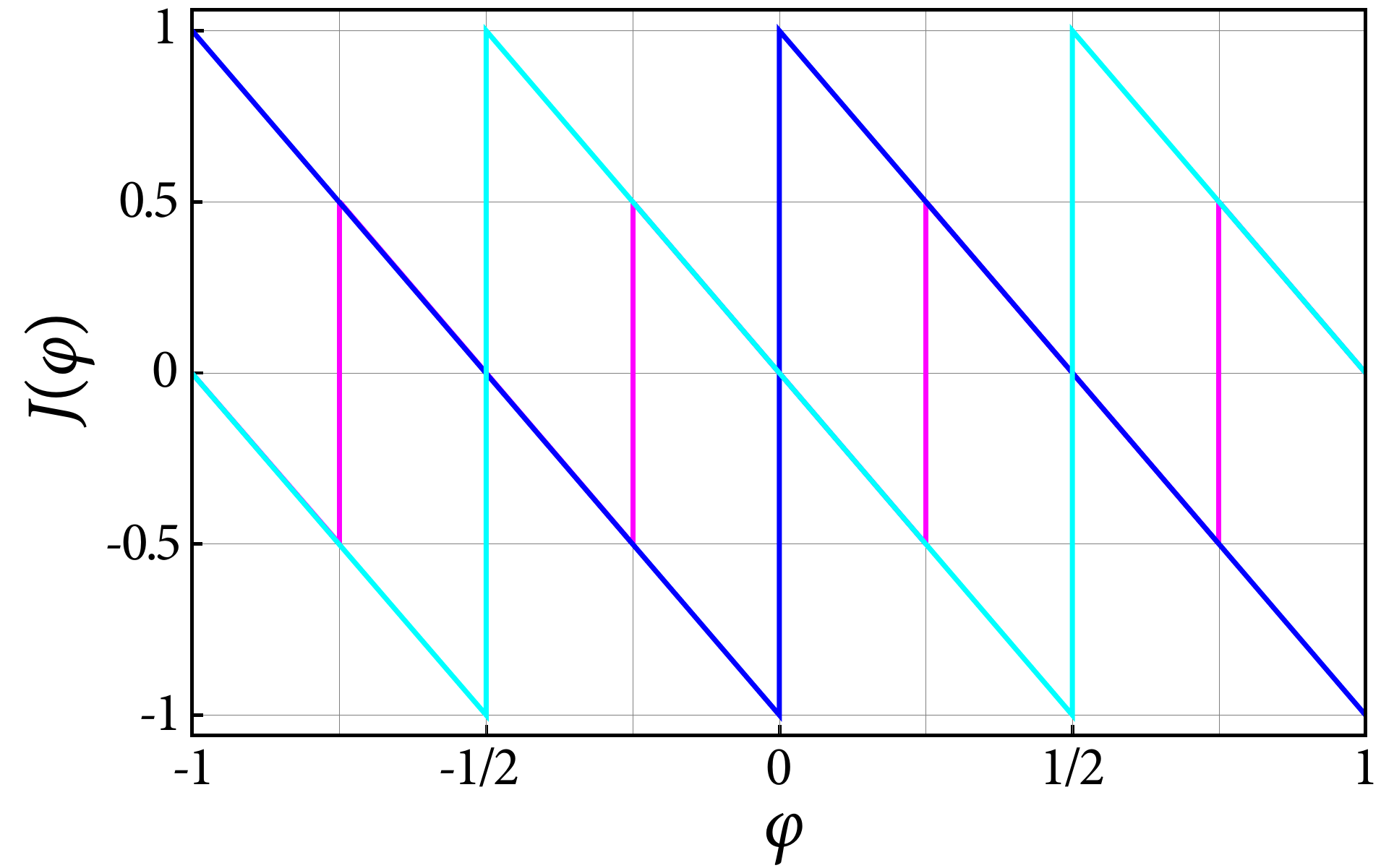}
\put(-3,59){(b)}
\end{overpic}\\[0mm]
\caption{(a) Total energy $E(\varphi)$ as a function of the magnetic flux $\varphi$. If the number of electrons on the ring $N$ is a multiple of four, then $E(\varphi)$ has minima at integer values of $\varphi$ (turquoise). If $N/4$ is an integer, then the minima are at half-integer values of $\varphi$ (blue). If $N/4$ is a half-integer, the parabolae are shifted by 1/2 (turquoise). The gray lines above the crossing points of the parabolae correspond to possible excited states. (b) Persistent current $J(\varphi)$ corresponding to the systems described in (a). The purple curve shows the current obtained for odd $N$.}
\label{fig:Sub_2}
\end{figure*}

The eigenenergies depend on the flux only in the combination $k-\varphi$, as is shown in figure~\ref{Figs0.10} for three different cases: (a) $N/4$ is an integer, (b) $N/4$ is a half integer, and (c) $N$ is an odd number. The $\varphi$ dependent shift in $\epsilon_k(\varphi)$ is known as the Doppler shift since it is proportional to the velocity of the corresponding electron. In all three cases, the spectrum has obviously the periodicity 1 with respect to $\varphi$. However, the flux values, for which an energy level crosses $E_{\rm F}$, are different. This number dependence, sometimes referred to as the ``parity effect\textquotedblright\ , is characteristic for discrete systems and not restricted to one dimension. It was discussed in detail in the context of the persistent current in metallic loops~\cite{landauer:85,cheung:88,oppen:91} and also in metallic nano clusters~\cite{mineev8}; it is also essential for the discussion of superconducting rings.

Physical quantities of the normal metal ring can be expressed through the thermal average $n_{s}(k)$ of the number of electrons with angular momentum $k$ and spin $s$: $n_{s}(k)=\langle c^\dag_{ks}c_{ks}\rangle=f(\epsilon_k(\varphi))$, with the Fermi distribution function $f(\epsilon)=1/(1+e^{\epsilon/k_{\rm B}T})$ for the temperature $T$. The groundstate is given by the minimum of the total energy $E$ of the system 
\begin{align}
E(\varphi)=\langle{\cal H}_0\rangle=\sum_{k,s}\epsilon_k(\varphi)n_{s}(k),
\label{mm6}
\end{align}
which is a piecewise quadratic function of the magnetic flux. The momentum distribution function $n_{s}(k)$ also depends on the magnetic flux only in the combination $k-\varphi$. The sum over $k$ in equation~(\ref{mm6}) directly renders the $\Phi_0$ flux periodicity of $E(\varphi)$. However, the position of the minima of $E(\varphi)$ depends on the highest occupied energy level and therefore also shows a parity effect (see figure~\ref{fig:Sub_2}).

The energy $E(\varphi)$ is maximal for those values of $\varphi$ where an energy level reaches $E_{\rm F}$ (red points) and has minima in between. If $N/4$ is an integer, then the minima of $E(\varphi)$ are at half-integer values of $\varphi$ (figure~\ref{fig:Sub_2}~(a), light blue curve), whereas if $N/4$ is a half integer, the minima are at integer values of $\varphi$ (figure~\ref{fig:Sub_2}~(a), dark blue curve). If $N$ is odd, two different (but physically equivalent) spectra for ${N}=4n\pm1$ are possible, and for both, two levels cross $E_{\rm F}$ in one flux period. This results in a superposition of the two previous cases and there are minima of $E(\varphi)$ for both integer and half-integer values of $\varphi$; $E(\varphi)$ is therefore $\Phi_0/2$ periodic. 

The normal persistent current $J(\varphi)=-(e/h)\,\partial E(\varphi)/\partial\varphi$ (see equation~\ref{bcs14.2} below) jumps whenever an energy level crosses $E_{\rm F}$, because the population of left and right circulating states changes abruptly [figure~\ref{fig:Sub_2}~(b)].  The occupied state closest to $E_{\rm F}$ contributes dominantly to the current, because all other contributions tend to almost cancel in pairs. The Doppler shift decreases with the ring radius like $1/R$ [c.f. equation~(\ref{bcs05})] and so does the persistent current.

\subsection{Superconducting state: Emergence of a new periodicity}\label{sec:em}

The theory of flux threaded superconducting loops was first derived by Byers and Yang~\cite{Byers}, Brenig~\cite{brenig:61}, and Onsager~\cite{onsager:61} on the basis of the BCS theory. They showed the  thermodynamic equivalence of the two superconducting states discussed above in the thermodynamic limit. However, in a strict thermodynamic limit the persistent (super-) current vanishes, and therefore a more precise statement is necessary with respect to the $\Phi_0/2$ periodicity of the supercurrent. Here we analyze the crossover from $\Phi_0$ periodicity in the normal metal loop to the $\Phi_0/2$ periodicity in the superconducting loop upon turning on the pairing interaction. The discussion of this crossover enables precise statements about the periodicity. 

For a one-dimensional superconducting loop (or any loop thinner than the penetration depth $\lambda$), finite currents flow throughout the superconductor. The magnetic flux is consequently not quantized, only the fluxoid $\Phi'=\Phi+(\Lambda/c)\oint{\rm d}{\bf r}\cdot{\bf J}({\bf r})$ is, which was introduced by F. London~\cite{London}. The flux $\Phi$ is the total flux threading the loop, including the current induced flux, and $\Lambda=4\pi\lambda^2/c^2$. In the absence of flux quantization, $\varphi=(e/hc)\,\varphi$ is a continuous variable also in a superconducting system with a characteristic periodicity in $\varphi$.

\begin{figure*}[t]	
\centering
\vspace{3mm}
\begin{overpic}
[width=0.95\columnwidth]{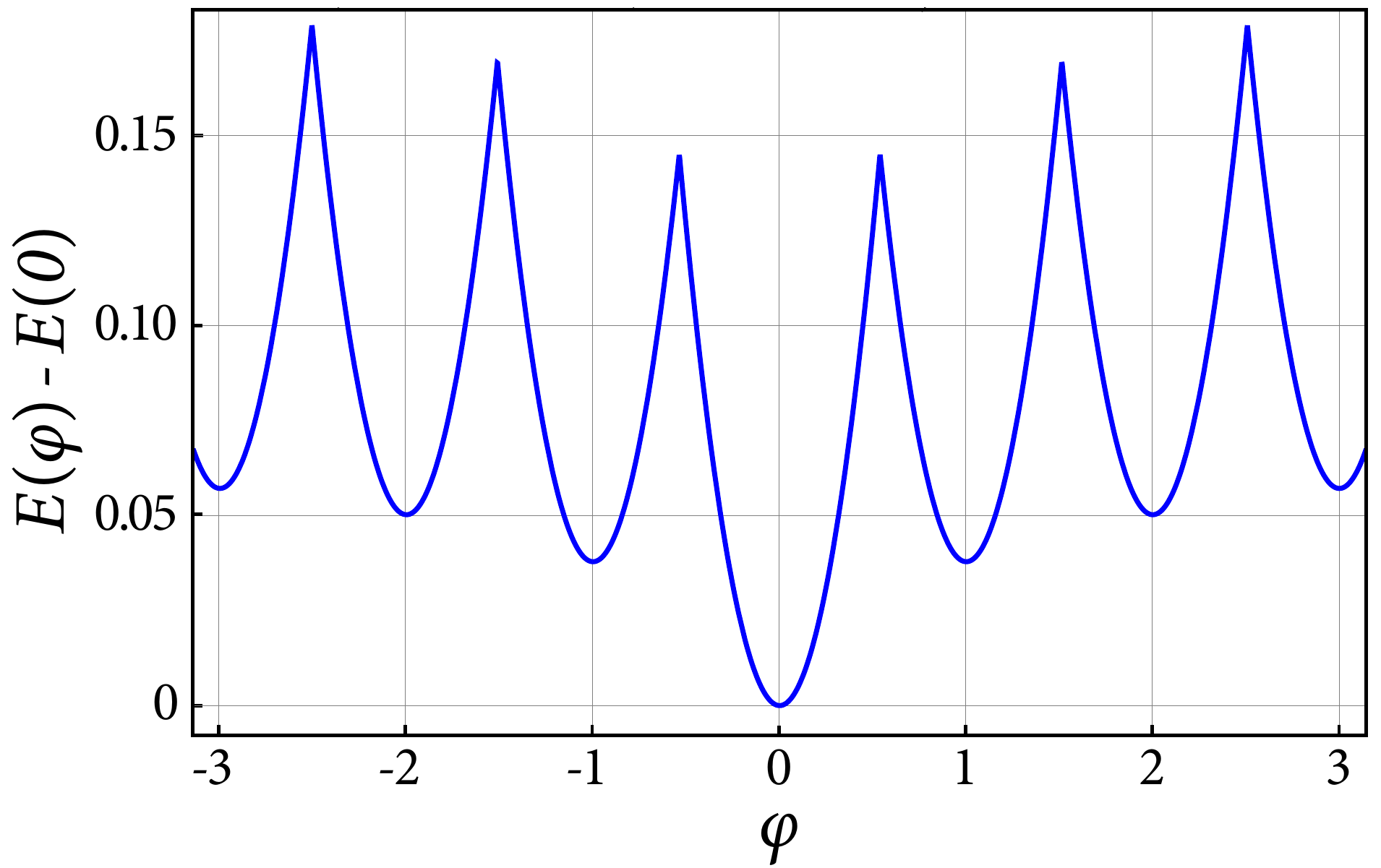}
\put(-3,59){(a)}
\end{overpic}
\hspace{6mm}
\begin{overpic}
[width=0.95\columnwidth]{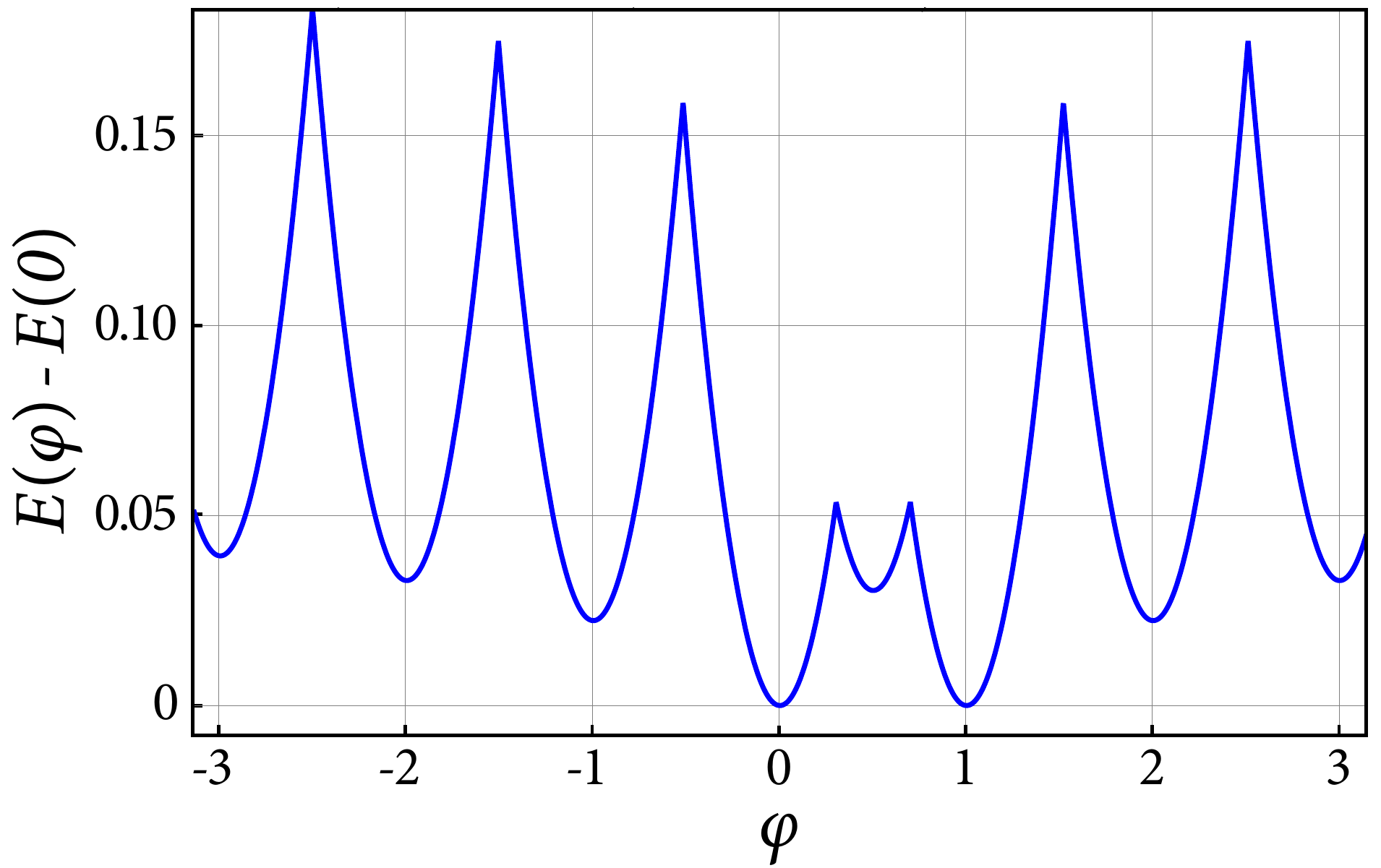}
\put(-3,59){(b)}
\end{overpic}\\[3mm]
\begin{overpic}
[width=0.95\columnwidth]{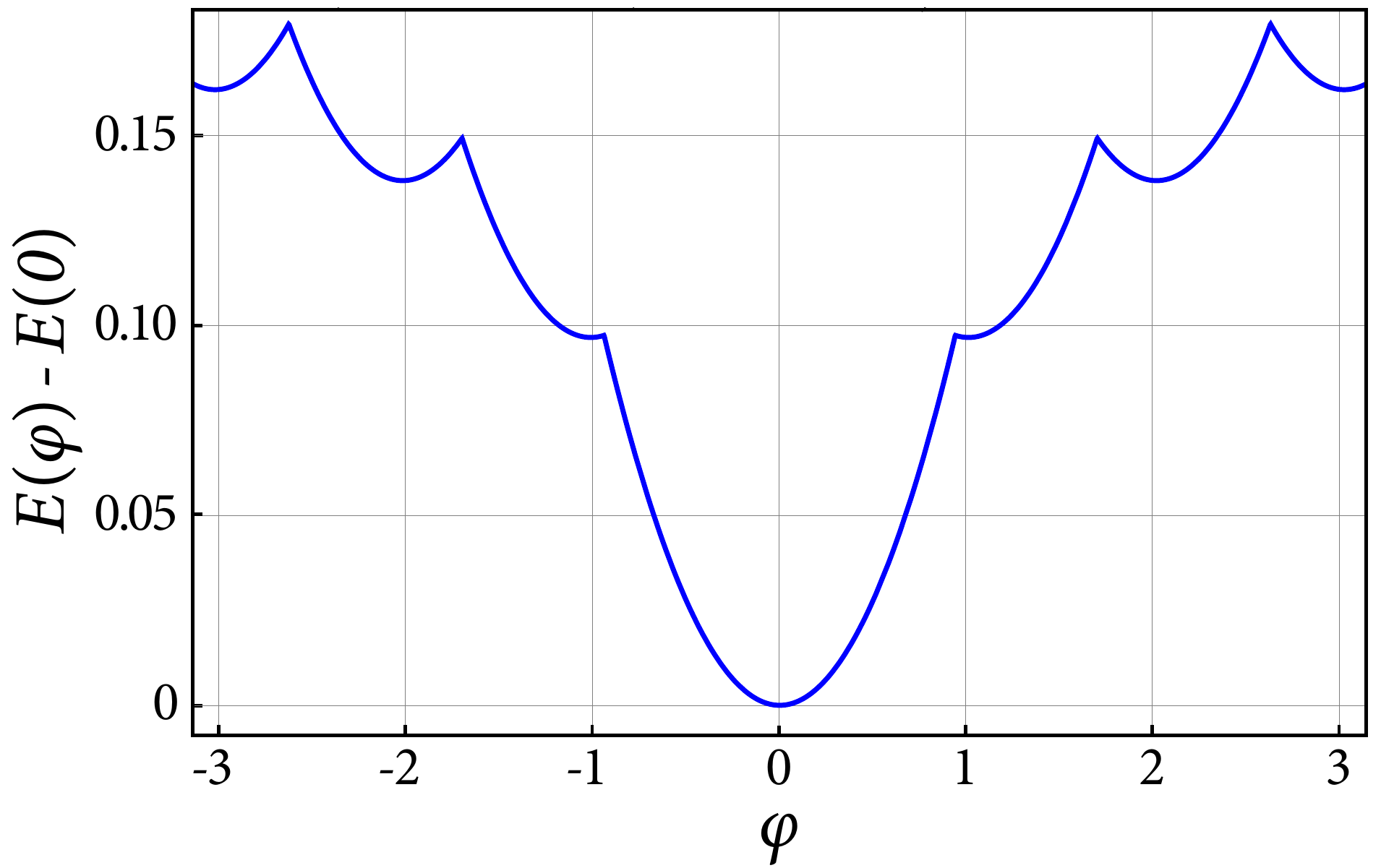}
\put(-3,59){(c)}
\end{overpic}
\hspace{6mm}
\begin{overpic}
[width=0.95\columnwidth]{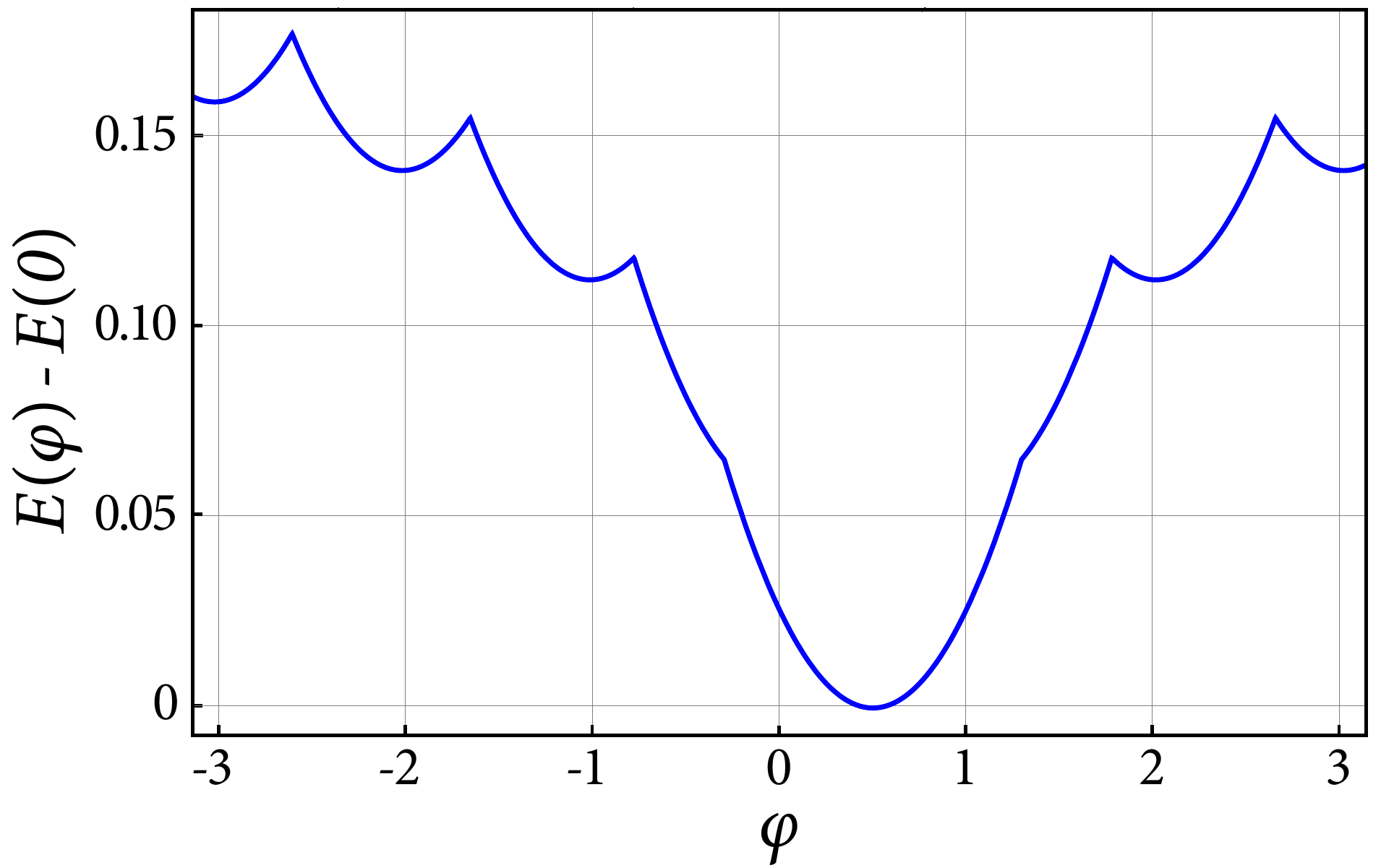}
\put(-3,59){(d)}
\end{overpic}\\[0mm]
\caption{Energy $E(\varphi)$ in the superconducting state with $q=0$ (a, c) and $q=1$ (b, d) for $N=50$. The upper panels (a, b) show the ``small gap\textquotedblright\  case with $\Delta=0.05\,t$ and the lower panels (c, d) the ``large gap\textquotedblright\  case with $\Delta=0.2\,t$.}
\label{fig:Sub_3}
\end{figure*}

In this section we focus on the emergence of a new periodicity when a superconducting order parameter arises. We therefore include an attractive on-site interaction of the general form
\footnote{In the literature the symmetric Hamiltonian $\tilde{\cal H}={\cal H}_0+(V/2N^2)\,\sum_{k,k'}\sum_q c^\dag_{k+q/2\ua}c^\dag_{-k+q/2\da}c_{-k'+q/2\da}c_{k'+q/2\ua}$ is often used~\cite{mineev17}. $\tilde{\cal H}$ is naturally $hc/e$ periodic in $\varphi$, but it is not well defined, although it yields the same physical quantities as $\cal H$. The introduction of half-integer angular momenta in $\tilde{\cal H}$ leads to two different limits $\Delta\rightarrow0$ for even or odd $q$, corresponding to the two spectra for $N/2$ even or odd. Therefore the symmetric $\tilde{\cal H}$ is unsuitable for the discussion of the flux periodicity.}
\begin{align}
{\cal H}={\cal H}_0-\frac{V}{2N^2}\sum_{k,k'}\sum_q c^\dag_{k\ua}c^\dag_{-k+q\da}c_{-k'+q\da}c_{k'\ua},
\label{H0}
\end{align}
where $V>0$ is the interaction strength.

\begin{figure*}[t]	
\centering
\vspace{3mm}
\begin{overpic}
[width=0.95\columnwidth]{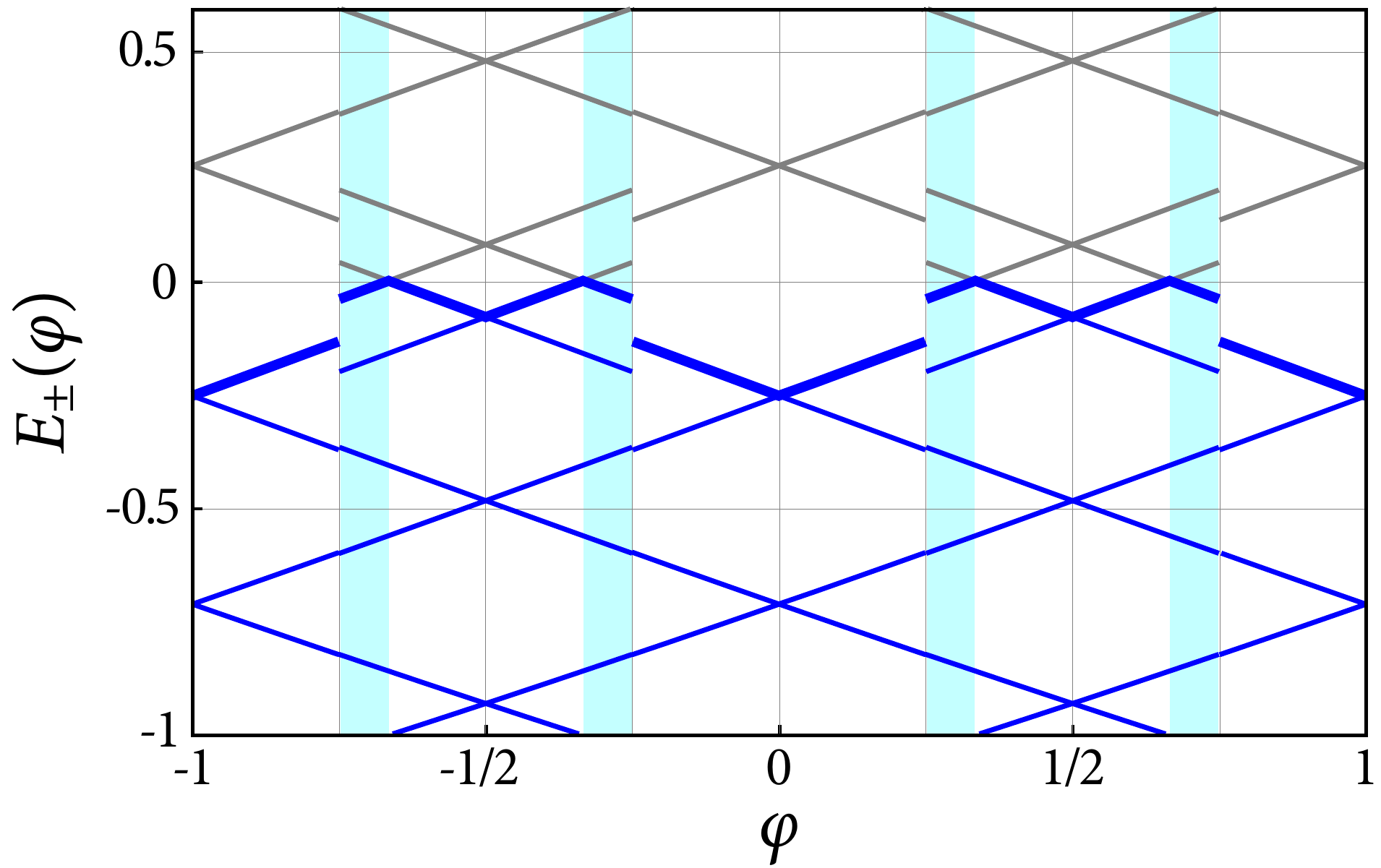}
\put(-3,59){(a)}
\end{overpic}
\hspace{6mm}
\begin{overpic}
[width=0.95\columnwidth]{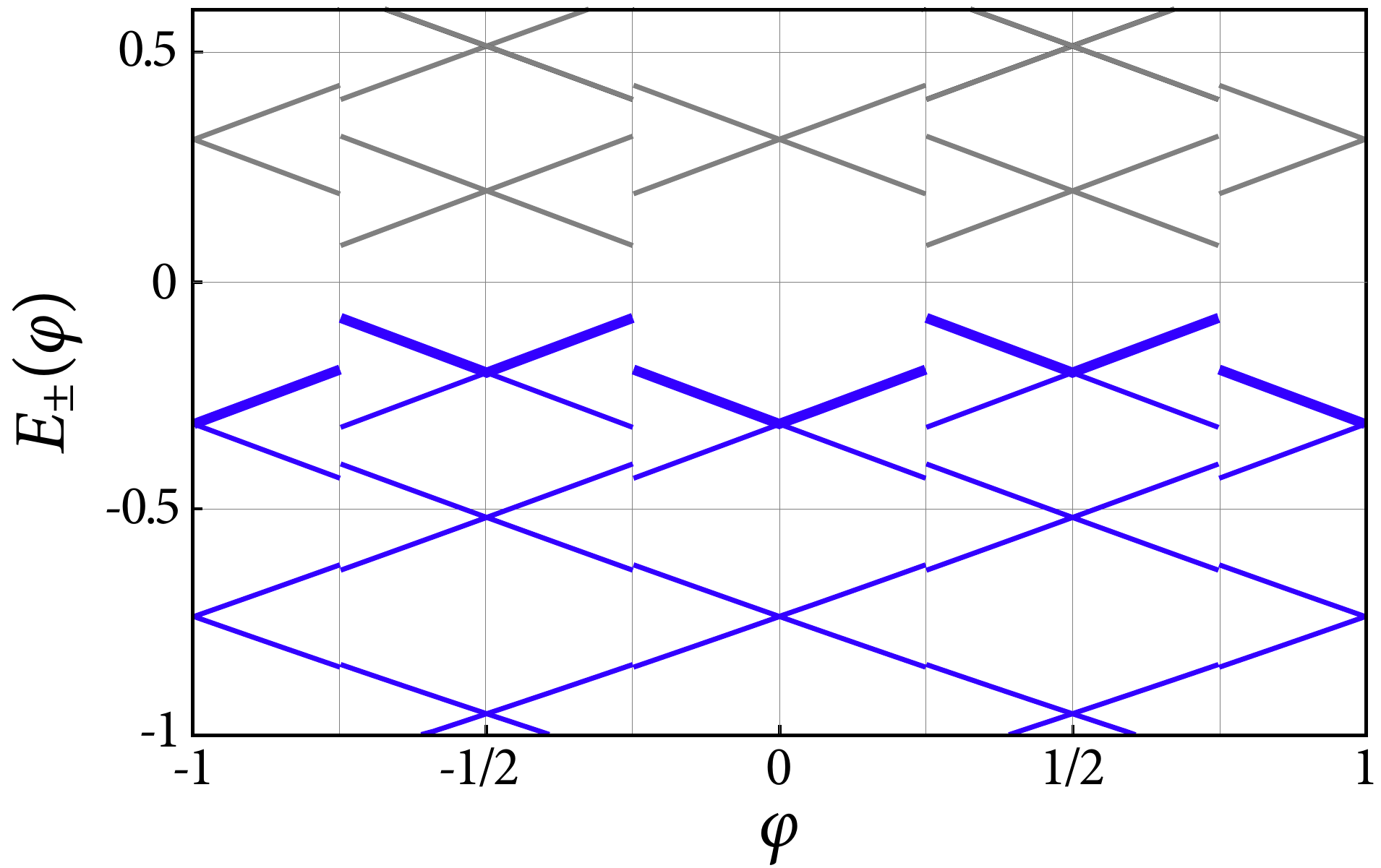}
\put(-3,59){(b)}
\end{overpic}\\[0mm]
\caption{Eigenenergies $E_\pm(k,\varphi)$ (\ref{s10}) as a function of flux $\varphi$ for $N=50$ and a fixed order parameter: (a) \textquotedblleft large gap\textquotedblright\  regime with $\Delta=0.2\,t$, (b) \textquotedblleft small gap\textquotedblright\ regime with $\Delta=0.05\,t$. Blue lines: occupied states, grey lines: unoccupied states. The bold line marks the highest 
occupied state for all $\varphi$. In the blue shaded areas in (a), the energy gap has closed due to the Doppler shift.}
\label{fig:Sub_4}
\end{figure*}

In BCS theory it is {\it assumed} that electron pairs have zero center-of-mass (angular) momentum, i.e., the pairs are condensed in a macroscopic quantum state with $q=0$, similar to a Bose-Einstein condensate of bosonic particles. In the case of a flux threaded ring, Byers and Yang~\cite{Byers}, Brenig~\cite{brenig:61}, and Onsager~\cite{onsager:61} showed that $q$ is generally finite and has to be chosen to minimize the kinetic energy of the Cooper pairs in the presence of a magnetic flux. Nevertheless it is still assumed that pairing occurs only for one specific angular momentum $q$. For conventional superconductors, this assumption is generally true, although for superconductors with gap nodes, the situation may be different, as was shown for $d$-wave pairing symmetry in reference~\cite{loder:10}. In this section, we use the assumption of condensation into a state with one selected angular momentum $q$ for all pairs, which allows us to write $\cal H$ in the decoupled form
\begin{multline}
{\cal H}={\cal H}_0+\sum_{k}\left[\Delta^{\!*}_q(\varphi)c_{-k+q\da}c_{k\ua}+\Delta_q(\varphi) c^\dag_{k\ua}c^\dag_{-k+q\da}\right]\\+\frac{\Delta_q^2(\varphi)}{V},
\label{bcs09}
\end{multline}
where the order parameter is defined as $\Delta_q(\varphi)=(V/2)\,\sum_k\langle c_{k\ua}c_{k\da}\rangle$. The mean-field Hamiltonian~(\ref{bcs09}) is diagonalized with the standard Bogoliubov transformation
\begin{align}
c_{k\ua}&=u(k)a_{k+}+v(k)a^\dag_{k-},\\c_{-k+q\da}&=u(k)a^\dag_{k-}-v(k)a_{k+}
\label{H2}
\end{align}
with the coherence factors
\begin{align}
u^2(k)&=\left(1+\frac{\epsilon(k,\varphi)}{E(k,\varphi)}\right),\\
v^2(k)&=\left(1-\frac{\epsilon(k,\varphi)}{E(k,\varphi)}\right),
\label{bcs14.1}
\end{align}
which depend on $\varphi$ and $q$ through
$E(k,\varphi)=\sqrt{\Delta_q^2+\epsilon^2(k,\varphi)}$ and $\epsilon(k,\varphi)=[\epsilon_{k}(\varphi)+\epsilon_{-k+q}(\varphi)]/2$.
The energy spectrum splits into the two branches
\begin{align}
E_{\pm}(k,\varphi)=\frac{\epsilon_{k}(\varphi)-\epsilon_{-k+q}(\varphi)}{2}\pm \sqrt{
\Delta_q^2+\epsilon^2(k,\varphi)},
\label{s10}
\end{align}
where the Doppler shift term arises from the different energies of the two paired states with momenta $k$ and $-k+q$. The order parameter $\Delta_q(\varphi)$ is determined self-consistently from
\begin{align}
\frac{1}{N}\sum_k\frac{f(E_-(k,\varphi))-f(E_+(k,\varphi))}{2\sqrt{\Delta_q(\varphi)^2+
\epsilon^2(k,\varphi)}}=\frac{1}{V}.
\label{s9.10}
\end{align}

For the discussion of the periodicity of this system, we first disregard the self-consistency condition for the order parameter and set $\Delta_q(\varphi)\equiv\Delta$ to be constant. Importantly, while ${\cal H}_0$ is strictly $\Phi_0$ periodic, $\cal H$ is {\it not} periodic in $\varphi$ if $\Delta>0$. The question of periodicity is therefore: {\it which} periodicity is restored by minimizing $E(\varphi)=\langle{\cal H}\rangle$ with respect to $q$ and how is this achieved?

Figure~\ref{fig:Sub_3} shows $E(\varphi)$ for two different values of $\Delta$ for $q=0$ and $q=1$. For small $\Delta$, $E(\varphi)$ is still a series of parabolae with minima at integer values of $\varphi$, but the degeneracy of the minima is lifted [figure~\ref{fig:Sub_3}~(a) and~(b)]. For even $q$, the lowest energy minimum is the one at $\varphi=q/2$, whereas for odd $q$, one new minimum at $\varphi=q/2$ emerges. If $\Delta$ exceeds a certain threshold $\Delta_{\rm c}$, this new odd $q$ minimum becomes deeper than the neighboring ones [figure~\ref{fig:Sub_3}~(d)]. We have thus identified the second class of states with minima in $E(\varphi)$ at half-integer flux values anticipated above and we find that the even and odd $q$ minima become equal if $\Delta$ becomes large compared to $\Delta_{\rm c}$, a ring size dependent value which we will determine below. It is to be understood that the energies $E(\varphi)$ in figure~\ref{fig:Sub_3} are not periodic in $\varphi$ because the $q$-values are fixed, either to $q=0$ in (a, c) or to $q=1$ in (b, d). In loops thicker than the penetration depth, screening currents drive the system always into an energy minimum. In this case, the flux is then quantized in units of $\Phi_0/2$.

\begin{figure*}[t]	
\centering
\vspace{3mm}
\begin{overpic}
[width=0.95\columnwidth]{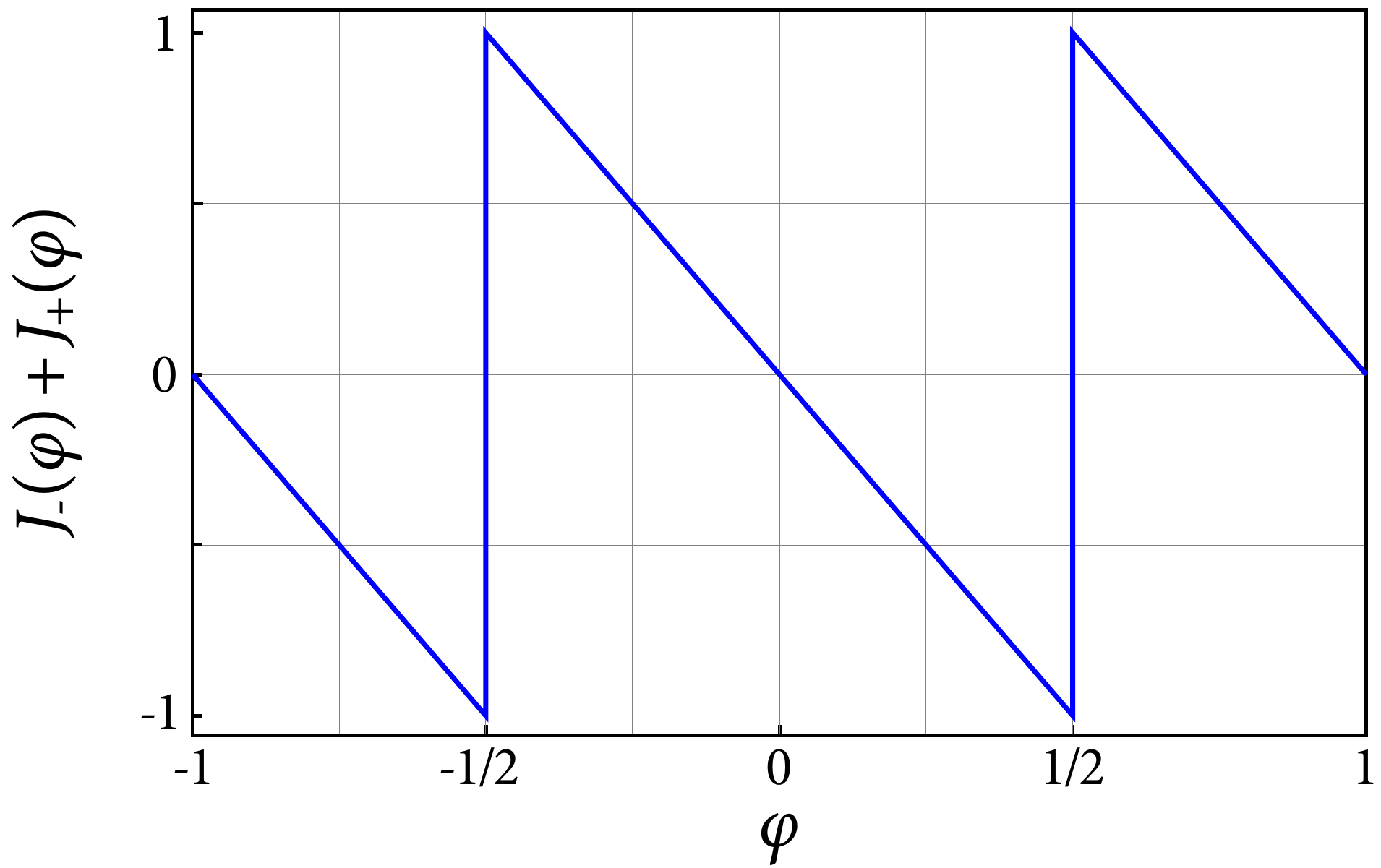}
\put(-3,59){(a)}
\end{overpic}
\hspace{6mm}
\begin{overpic}
[width=0.95\columnwidth]{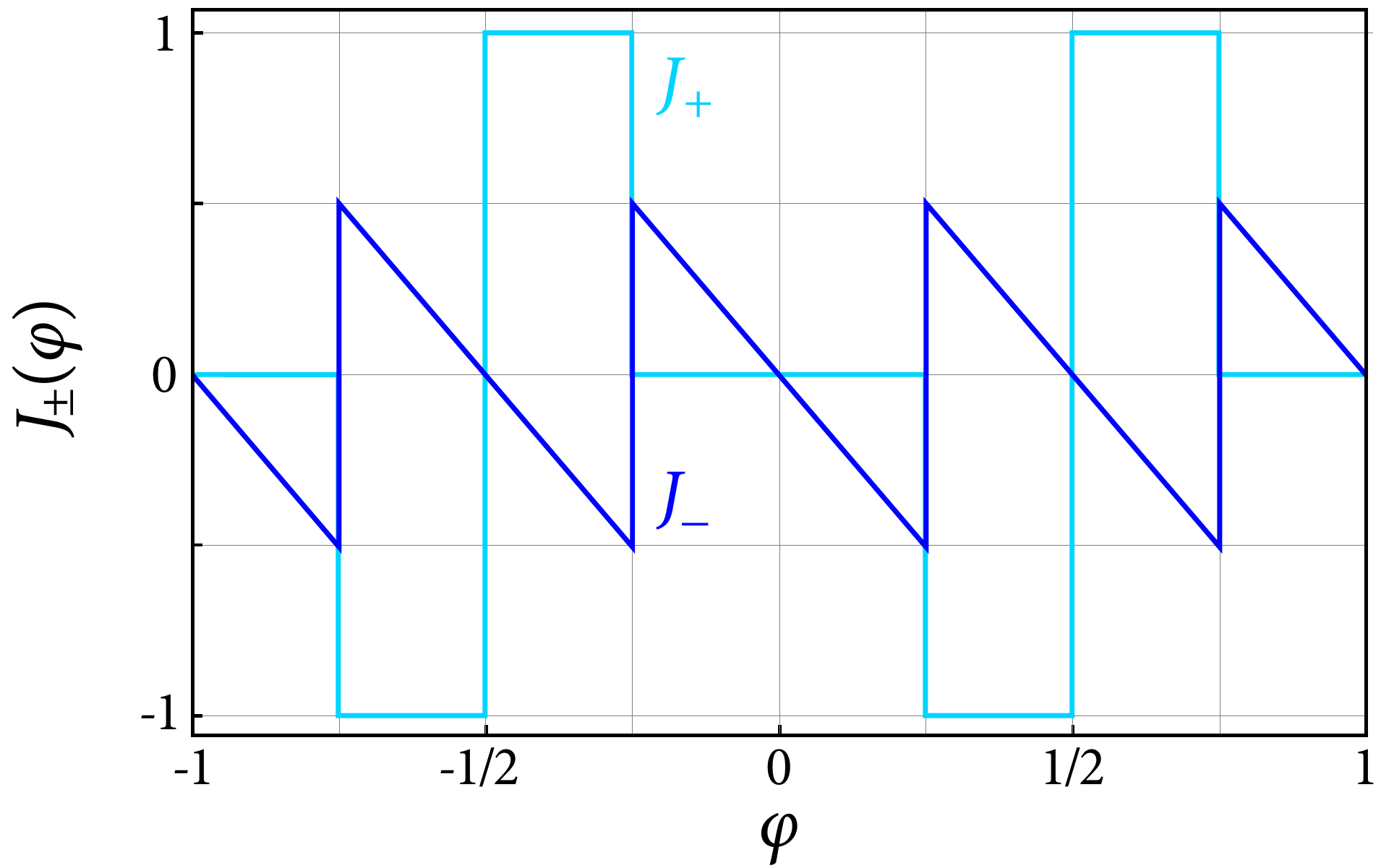}
\put(-3,59){(b)}
\end{overpic}\\[0mm]
\caption{(a) The $\Phi_0$ periodic persistent current $J(\varphi)$ (in units of $t/\Phi_0$) in the normal state. (b) The contribution $J_-(\varphi)$ (dark blue) is $\Phi_0/2$ periodic and identical in the normal and the superconducting state. The $\Phi_0$ periodicity in the \textquotedblleft small gap\textquotedblright\ regime is entirely due to $J_-(\varphi)$ (light blue), which vanishes in the  \textquotedblleft large gap\textquotedblright\ regime.}
\label{fig:Sub_5}
\end{figure*}

\begin{figure*}[t]
\centering
\includegraphics[height=41mm]{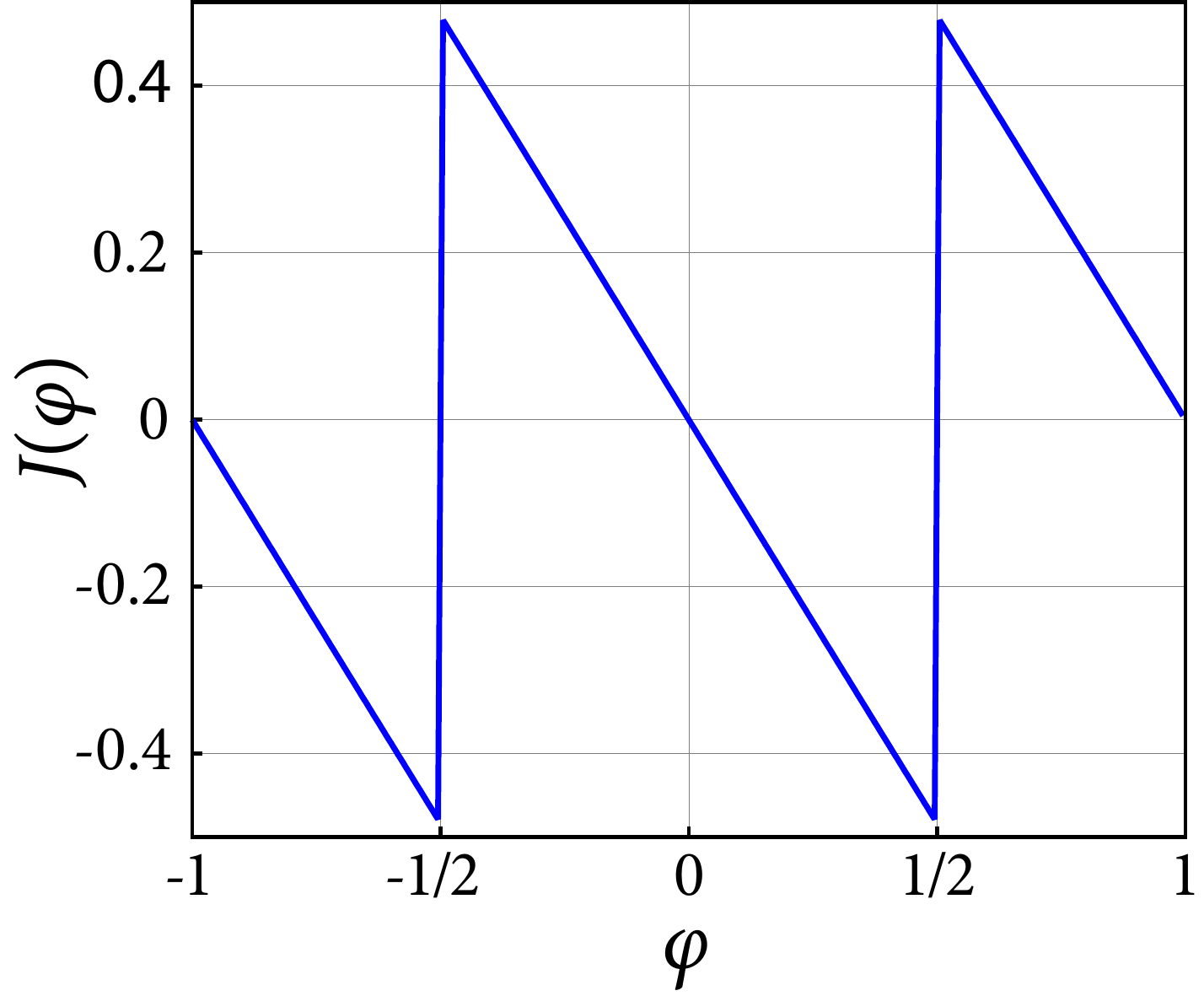}\!\!
\includegraphics[height=41mm]{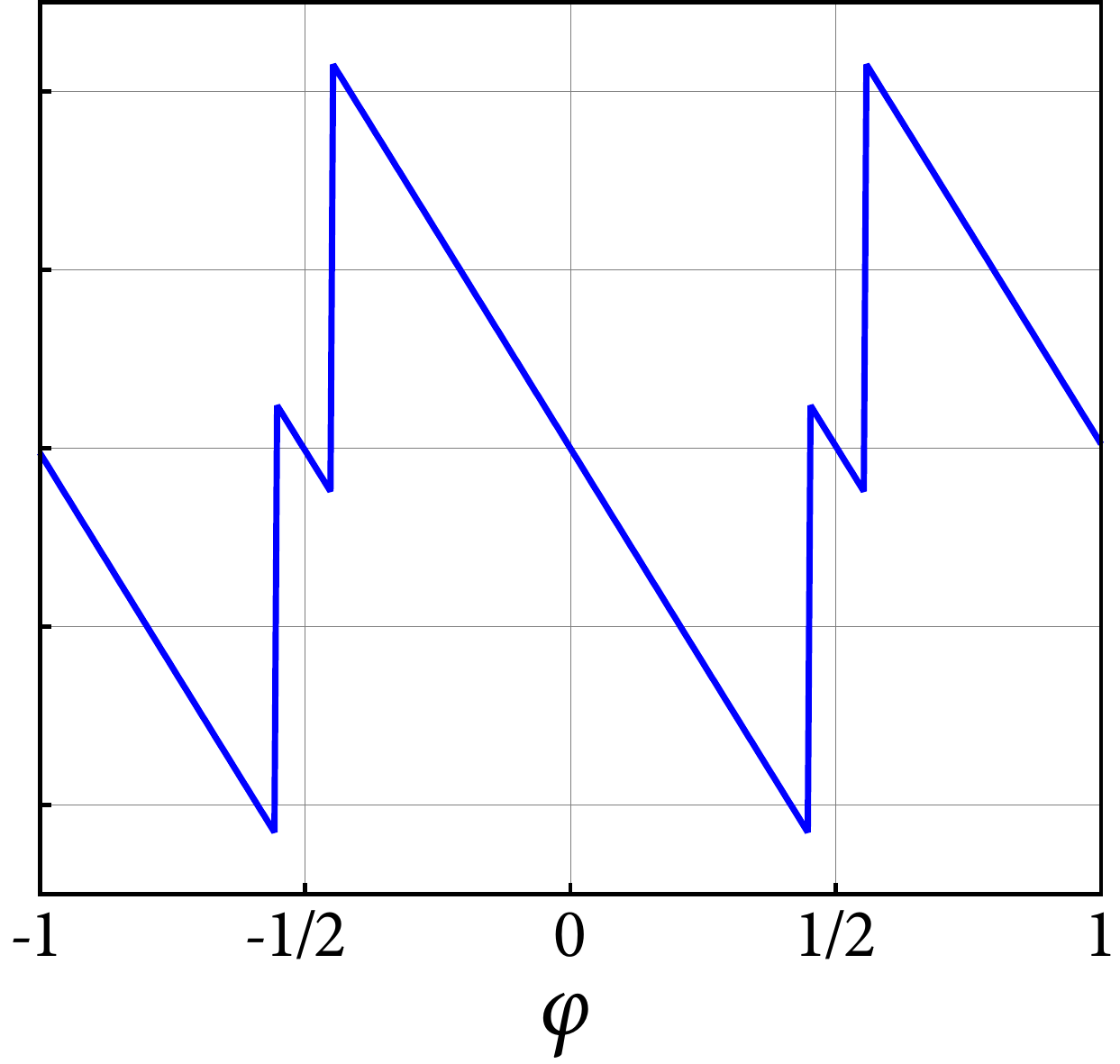}\!\!
\includegraphics[height=41mm]{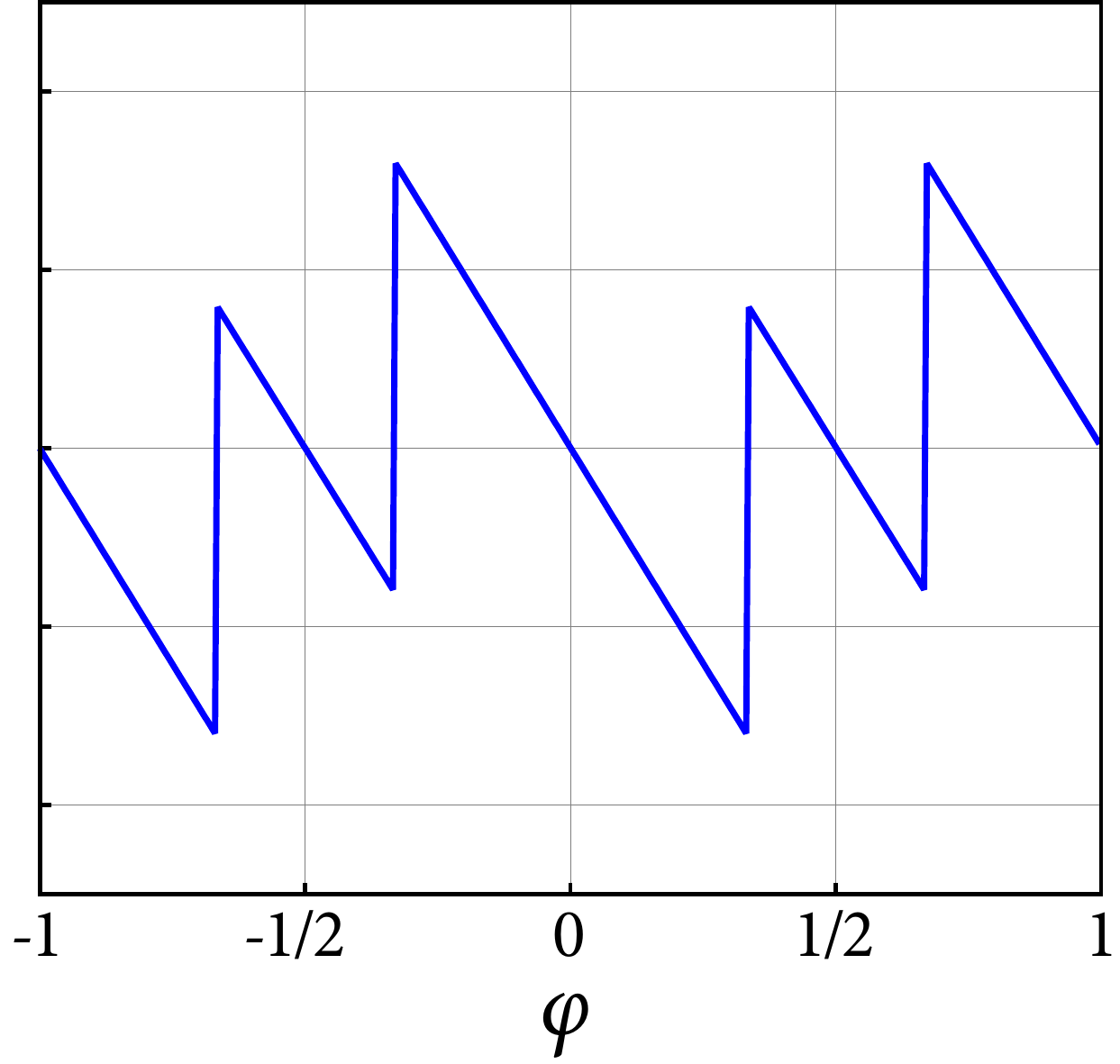}\!\!
\includegraphics[height=41mm]{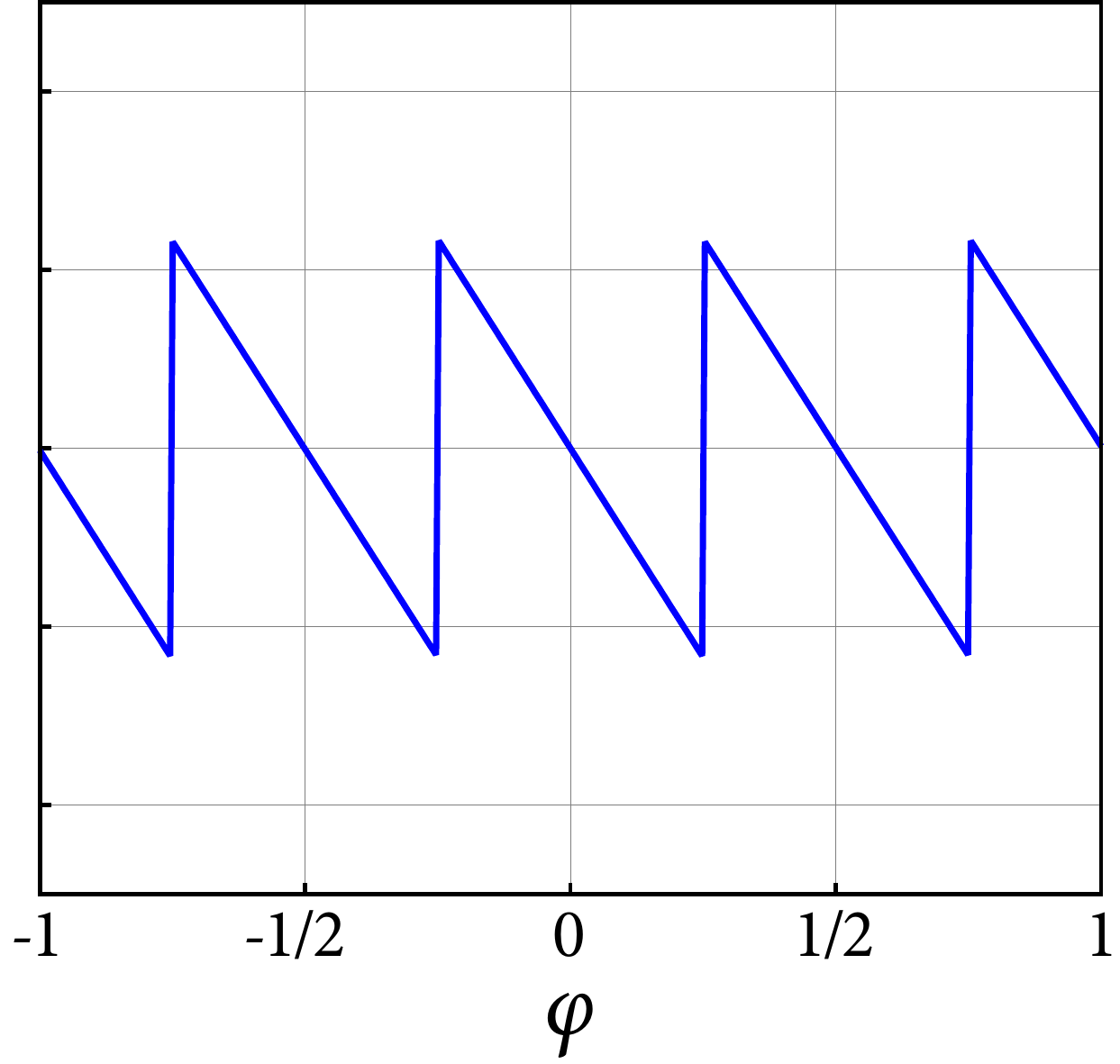}
\caption{
Crossover from the $\Phi_0$-periodic normal persistent current to the $\Phi_0/2$-periodic 
supercurrent in a ring with $N=26$ at $T=0$. $J(\varphi)$ is in units of $t/\Phi_0$. For this ring size $\Delta_{\rm c}
\approx0.24\,t$. The discontinuities occur where the $\varphi$-derivative of the 
highest occupied state energy changes sign. From left to right: $\Delta=0,\ 0.08\,t,\ 0.16\,t,\ 0.24\,t$.}
\label{Figs3}
\end{figure*}

Let us for the moment assume that the flux value, at which the energy minimizing $q$ changes from one integer to the next, is well approximated by the half way between two minima: $q=\text{floor}(2\varphi+1/2)$ ($\text{floor}(x)$ is the largest integer smaller than $x$, e.g., $\varphi=0\rightarrow q=0$; $\varphi=1/4\rightarrow q=1$; $\varphi=3/4\rightarrow q=2$). Small deviations from these values will be discussed for the self-consistent solution in section~\ref{sec:self}. The energy spectrum is then $\Phi_0$ periodic, but discontinuous at the flux values where $q$ changes, as shown in figure~\ref{fig:Sub_4}. Clearly distinguishable are now the ``small gap\textquotedblright\  (a) and the ``large gap\textquotedblright\  (b) regime: $\Delta_{\rm c}$ represents the maximum of the flux-induced shift of the energy levels close to $E_{\rm F}$, before $q$ changes. If $\Delta<\Delta_{\rm c}$, the energy gap closes at certain values of $\varphi$, whereas if $\Delta>\Delta_{\rm c}$, an energy gap persists for all $\varphi$.

Although the spectra are $\Phi_0$ periodic both in figure~\ref{fig:Sub_4}~(a) and~(b), the closing of the gap in the ``small gap\textquotedblright\  regime has significant effects on the periodicity of physical quantities like $E(\varphi)$. Even more prominent is the periodicity crossover for the persistent current in the ring. The supercurrent is given by $J(\varphi)=J_+(\varphi)+J_-(\varphi)=(e/h)\partial E(\varphi)/\partial\varphi$, where
\begin{align}
J_\pm(\varphi)=\frac{e}{hc}\sum_{k}\frac{\partial\epsilon_{k}(\varphi)}{\partial k}n_\pm(k)
\label{bcs14.2}
\end{align}
with $n_+(k)=u^2(k)f(E_+(k,\varphi))$ and $n_-(k)=v^2(k)f(E_-(k,\varphi))$.
$J_+(\varphi)$ and $J_-(\varphi)$, as well as $J(\varphi)$, are plotted in figure~\ref{fig:Sub_5}. The contribution $J_-(\varphi)$ forms a $\Phi_0/2$ periodic saw-tooth pattern, both in the normal and in the superconducting state. The $\Phi_0$ periodic part in the normal state is contained exclusively in $J_+(\varphi)$. A flux window where $E_+(k,\varphi)$ is partially occupied exists in each $q$ sector when the energy gap has closed [shaded blue areas in figure~\ref{fig:Sub_4}~(a)]. These windows decrease for increasing $\Delta$ until $J_+(\varphi)$ vanishes for $\Delta=\Delta_{\rm c}$. In the \textquotedblleft large gap" regime, the supercurrent is carried entirely by $J_-(\varphi)$ and is therefore $\Phi_0/2$ periodic and essentially independent of $\Delta$. The discontinuities in $J_-(\varphi)$ are not caused by energy levels crossing $E_{\rm F}$, but by the reconstruction of the condensate when the pair momentum $q$ changes to the next integer at the flux values $\varphi=(2n-1)/4$. Figure~\ref{Figs3} shows the periodicity crossover of the persistent current in four different steps from the $\Phi_0$ periodic normal current to the $\Phi_0/2$ periodic supercurrent.

Further insight into the current 
periodicity is obtained by analyzing $\Delta_{\rm c}$. Close to $E_{\rm F}$, the maximum 
energy shift is $\epsilon_{\rm D}=at/2R$, and the condition for a direct energy gap (or 
$E_+(k,\varphi)>0$ for all $k$, $\varphi$) and a $\Phi_0/2$-periodic current pattern is 
therefore $\Delta>\Delta_{\rm c}=\epsilon_{\rm D}$. The corresponding critical ring radius 
is $R_{\rm c}=at/2\Delta$.
It is instructive to compare $R_{\rm c}$ with the BCS coherence length 
$\xi_0=\hslash v_{\rm F}/\pi\Delta$, where $v_{\rm F}$ is the Fermi velocity and $\Delta$ 
the BCS order parameter at $T=0$. On the lattice we identify $v_{\rm F}=\hslash k_{\rm F}/m$ 
with $k_{\rm F}=\pi/2a$ and $m=\hslash^2/2a^2t$, and obtain $\xi_0=at/\Delta$ and thus $2R_{\rm c}=\xi_0$. This 
signifies that the current response of a superconducting ring with a diameter smaller than 
the coherence length, is generally $\Phi_0$ periodic \cite{loder:08}. In these rings the Cooper-pair wavefunction is delocalized around the ring.

We have hereby identified the basic mechanism underlying the crossover from $\Phi_0$ periodicity in the normal state to $\Phi_0/2$ periodicity in the superconducting state. It is the crossing of $E_{\rm F}$ of energy levels as a function of the magnetic flux that leads to kinks in the energy and to discontinuities in the supercurrent (or the persistent current in the normal state). If the superconducting gap is large enough to prevent all energy levels from crossing the Fermi energy, the kinks and jumps occur only where the pair momentum $q$ of the groundstate changes. The latter is true, if the ring diameter is larger than the coherence length $\xi_0$ of the superconductor.

To conclude the discussion of the supercurrent we mention an issue raised by Little and Parks~\cite{Parks}. A simple theoretical model to predict the amplitude of the oscillations of $T_{\rm c}$ is the following: For all non-integer or non-half-integer values of $\varphi$, there is a persistent current $J(\varphi)$ circulating in the cylinder. The kinetic energy $E_{\rm kin}(\varphi)$ associated with this current is proportional to $J^2(\varphi)$, as is the energy $E(\varphi)$ in figure~\ref{fig:Sub_2}~(a). It is therefore suggestive to subtract  $E_{\rm kin}(\varphi)$ from the condensation energy of the superconducting state and deduce the oscillations of $T_{\rm c}$ from those of $E(\varphi)$. This was done in a first approach by Little and Parks~\cite{Little}, by Tinkham~\cite{Tinkham}, and by Douglass~\cite{douglass} within a Ginzburg-Landau ansatz, yet it was later shown to be incorrect by Parks and Little in a subsequent article~\cite{Parks}. They wrote that \textquotedblleft the microscopic theory [i.e., the BCS theory] shows that it is {\it not the kinetic energy of the pairs} which raises the free energy of the superconducting phase ..., but rather it is due to {\it the difference in the energy} of the two members of the pairs", i.e., $\epsilon_k(\varphi)-\epsilon_{-k+q}(\varphi)$. It is remarkable that the results of Tinkham and Douglass are nevertheless identical to the microscopic result~\cite{peshkin:63}. The notion whether it is the kinetic energy of the screening current that causes the oscillations, or rather an internal cost in condensation energy in the presence of a finite flux, is important insofar as it provides an explanation for an intriguing problem: In the same way as the pairing of electrons leads to a reduction of the fundamental flux period $\Phi_0$ to $\Phi_0/2$, the pairing of pairs to quartets would lead to the quarter-period $\Phi_0/4$. Then the saw-tooth pattern of the supercurrent becomes $\Phi_0/4$ periodic and the maximum current is only half the value for unpaired Cooper pairs. If the oscillation in $E(\varphi)$ were due to the kinetic energy of the pairs, then the formation of quartets and the $\Phi_0/4$ periodicity would be energetically favorable. The fact that it is $\Phi_0/2$ periodic illustrates the remark by Parks and Little.

\subsection{Multi channels and self consistency}\label{sec:self}

\begin{figure}[t]	
\centering
\includegraphics[width=0.75\columnwidth]{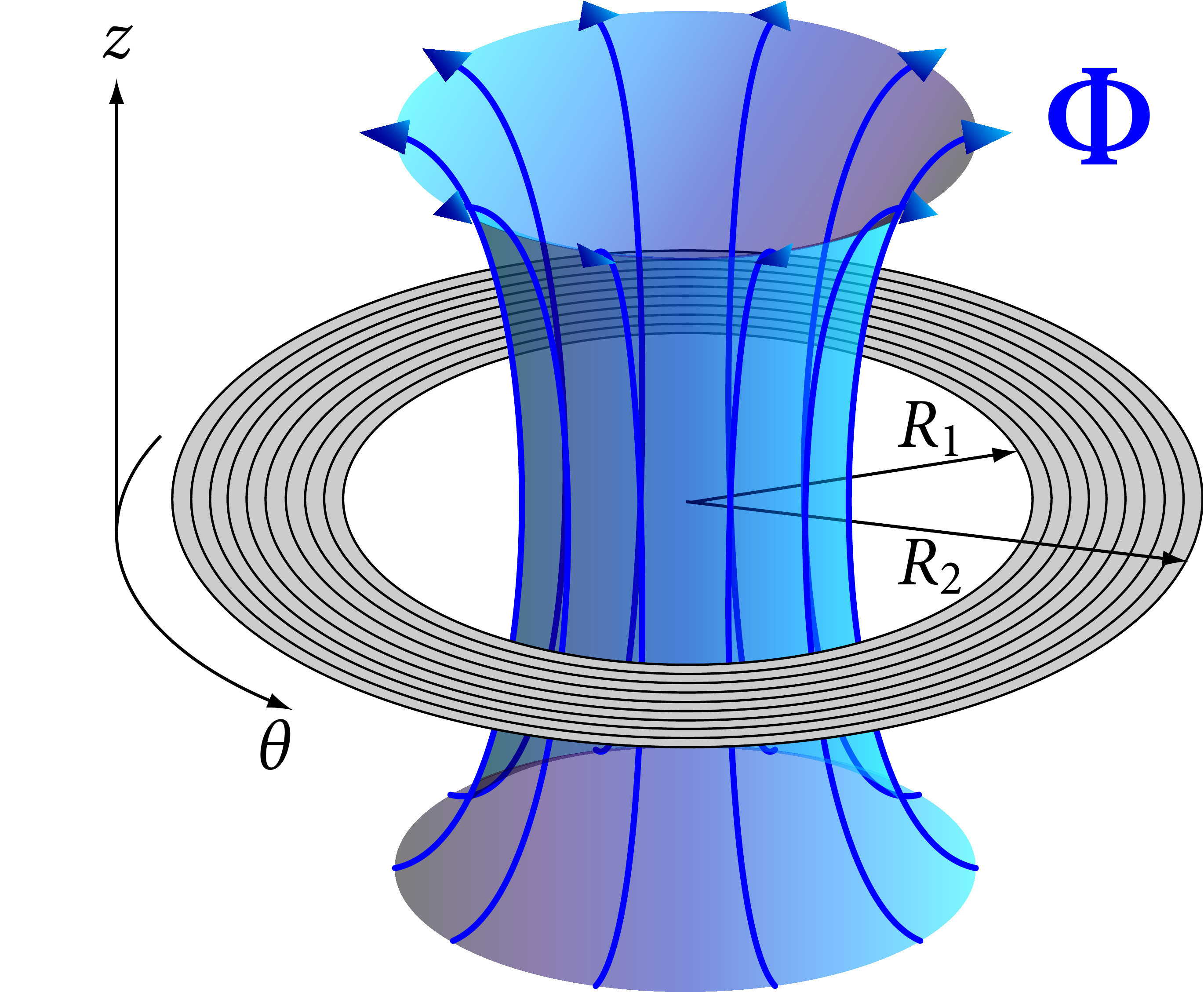}
\vspace{3mm}
\caption{Flux threaded annulus with inner radius $R_1$ and outer radius $R_2$. For a magnetic flux threading the interior of the annulus, the radial part of the Bogoliubov - de Gennes equations is solved numerically with a discretized radial coordinate.}
\label{fig:Sub_6}
\end{figure}

Although the one-dimensional ring discussed above comprises all the qualitative features of the periodicity crossover at $T=0$ upon entering the superconducting state, some additional issues need to be considered. First, the spectrum of a one dimensional ring is special insofar as only two energy levels exist that cross the Fermi energy in one flux period. This situation is ideal to investigate persistent currents, because there are maximally two jumps in one period. In an extended loop all radial channels contribute at the Fermi energy and have to be taken into account. Second, the self-consistency condition of the superconducting order parameter leads to corrections of the results obtained above, and third, the periodicity crossover upon entering the superconducting state by cooling through the transition temperature $T_{\rm c}$ is somewhat different from the $T=0$ crossover. These points are addressed in this section.

Here we extend the ring to an annulus with an inner radius $R_1$ and an outer radius $R_2$, as shown in figure~\ref{fig:Sub_6}. For such an annulus, we choose a continuum approach on the basis of the Bogoliubov - de Gennes (BdG) equations, for which no complications arise from the parity effect. 
For spin singlet pairing the BdG equations are~\cite{degennes}
\begin{align}
\begin{split}
E_{\bf n}u_{\bf n}({\bf r})&=\left[\frac{1}{2m}\left(i\hslash\bm\nabla+\frac{e}{c}{\bf A({\bf r})}\right)^2-\mu\right]u_{\bf n}({\bf r})\\&\makebox[4cm]{}+\Delta({\bf r})\,v_{\bf n}({\bf r}),
\end{split}\\[3mm]
\begin{split}
E_{\bf n}v_{\bf n}({\bf r})&=-\left[\frac{1}{2m}\left(i\hslash\bm\nabla-\frac{e}{c}{\bf A({\bf r})}\right)^2-\mu\right]v_{\bf n}({\bf r})\\&\makebox[4cm]{}+\Delta^{\!*}({\bf r})u_{\bf n}({\bf r}),
\end{split}
\label{1}
\end{align}
with the self-consistency condition (gap equation) for the order parameter $\Delta({\bf r})$:
\begin{align}
\begin{split}
\Delta({\bf r})=V\sum_{\bf n}u_{\bf n}({\bf r})v^*_{\bf n}({\bf r})\tanh\left(\frac{E_{\bf n}}{2k_{\rm B}T}\right),
\end{split}
\label{34}
\end{align}
where $V$ is the local pairing potential. For an annulus of finite width we separate the angular part of the quasi-particle wavefunctions $u_{\bf n}({\bf r})$, $v_{\bf n}({\bf r})$ using polar coordinates ${\bf r}=(r,\theta)$ with the ansatz
\begin{align}
\begin{split}
u_{\bf n}(r,\theta)&=u_{\bf n}(r)e^{\frac{i}{2}(k+q)\theta},\\
v_{\bf n}(r,\theta)&=v_{\bf n}(r)e^{\frac{i}{2}(k-q)\theta},
\end{split}
\label{2}
\end{align}
where $k$ and $q$ are either both even or both odd integers. Thus  $\hslash k$ is the angular momentum as for the one dimensional ring and ${\bf n}=(k,\rho)$ with a radial quantum number $\rho$.
The order parameter factorizes into $\Delta(r,\theta)=\Delta(r)e^{iq\theta}$ where the radial component
\begin{align}
\Delta(r)=V_0\sum_{\bf n}u_{\bf n}(r)v^*_{\bf n}(r)\tanh\left(\frac{E_{\bf n}}{2k_{\rm B}T}\right)
\label{38}
\end{align}
is real. For a magnetic flux $\varphi$ threading the interior of the annulus we choose the vector potential ${\bf A}(r,\theta)={\bf e}_\theta\,\varphi/(2\pi r)$, where ${\bf e}_\theta$ is the azimuthal unit vector. With
\begin{align}
\left(-i\bm\nabla\pm\frac{\varphi}{r}{\bf e}_\theta\right)^2=-\frac{1}{r}\partial_r(r\partial_r)+\frac{1}{r^2}(-i\partial_\theta\pm\varphi)^2
\label{3}
\end{align}
the BdG equations therefore reduce to radial differential equations for $u_{\bf n}(r)$ and $v_{\bf n}(r)$:
\begin{align}
\begin{split}
E_{\bf n}\,u_{\bf n}(r)&=\!-\!\!\left[\frac{\hslash^2}{2m}\frac{\partial_r}{r}(r\partial_r)\!-\!\frac{\hslash^2l_u^2}{2mr^2}\!+\!\mu\right]\!u_{\bf n}(r)\!+\Delta(r)v_{\bf n}(r),\\
E_{\bf n}\,v_{\bf n}(r)&=\!\left[\frac{\hslash^2}{2m}\frac{\partial_r}{r}(r\partial_r)\!-\!\frac{\hslash^2l_v^2}{2mr^2}\!+\!\mu\right]\!v_{\bf n}(r)\!+\Delta(r)u_{\bf n}(r),
\end{split}
\label{4}
\end{align}
with the canonical angular momenta
\begin{align}
\hslash l_u&=\frac{\hslash}{2}(k+q-2\varphi),\\
\hslash l_v&=\frac{\hslash}{2}(k-q+2\varphi).
\label{5}
\end{align}
For integer and half-integer flux values, equations~(\ref{4}) can be solved analytically whereas for an arbitrary magnetic flux a numerical solution is required (for details, see reference~\cite{loder:08.2}). Within this procedure, the radial coordinate is discretized into $N_\perp$ values $r_n$ separated by the distance $a_\perp=(R_2-R_1)/N_\perp$.
The number $q$ plays the same role as in the previous section. Here we choose $q$ for each value of the flux to minimize the total energy of the system. The flux for which $q$ changes to the next integer can therefore deviate from the values $(2n-1)/4$, at which we fixed the jump to the next $q$ for the one-dimensional model. 

\begin{figure}[t]	
\centering
\vspace{3mm}
\begin{overpic}
[width=0.95\columnwidth]{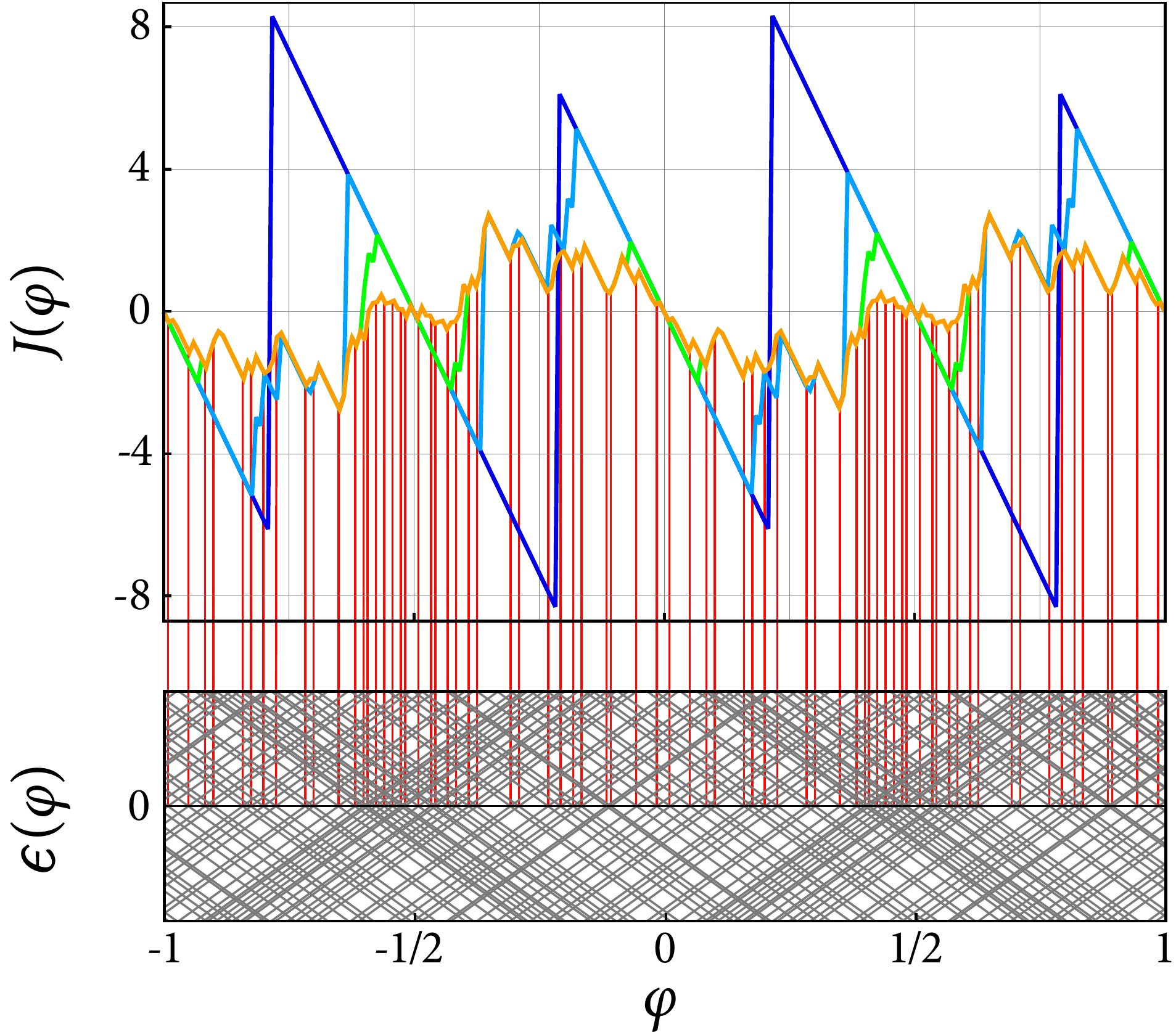}
\put(-3,85){(a)}
\end{overpic}
\vspace{6mm}\\
\begin{overpic}
[width=0.95\columnwidth]{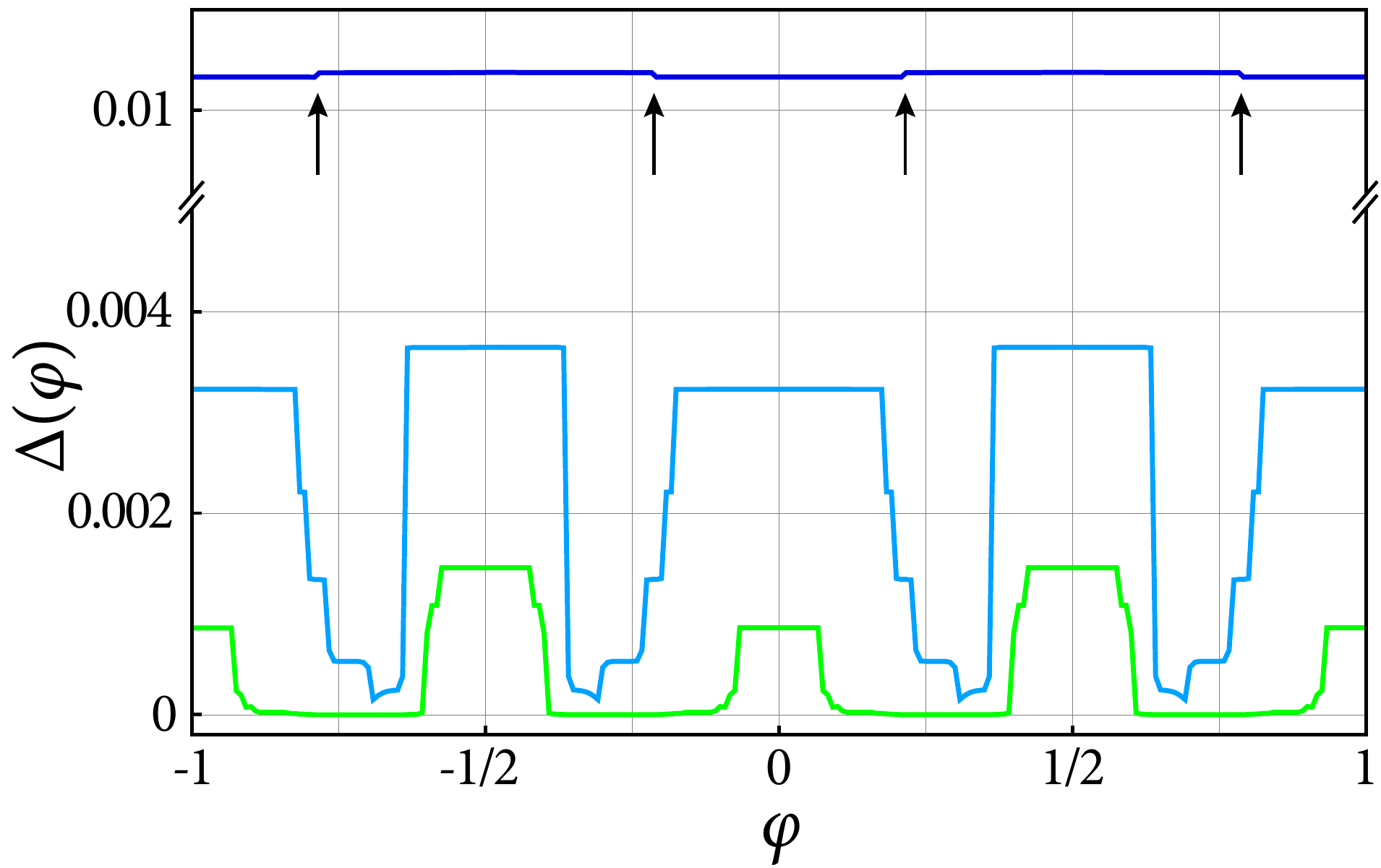}
\put(-3,59){(b)}
\end{overpic}\\[0mm]
\caption{Self-consistent calculations for a discretized annulus with an inner radius $R_1=100a_\perp$ and an outer radius $R_2=150a_\perp$. The (super-) current (a) $J(\varphi)$ jumps whenever an energy level crosses $E_{\rm F}$. The energy levels $\epsilon(\varphi)$ of the normal state are indicated by the grey lines.  (b) displays the self-consistent order parameter $\Delta$ as a function of $\varphi$. The lines correspond to the pairing interaction $V=0$ (orange), $V=0.28\,t$ (green), $V=0.32\,t$ (turquoise), and $V=0.38\,t$ (blue). The black arrows mark the positions of the $q$-jump for $V=0.38\,t$. Here the energy units are $t=\hslash^2/2m_ea_\perp$.}
\label{fig:Sub_7}
\end{figure}

In the normal state ($\Delta=0$), the number of eigenstates sufficiently close to $E_{\rm F}$ which may cross $E_{\rm F}$ as a function of $\varphi$ is controlled by the average charge density $n$ and is approximately $(R_2-R_1)/a_\perp$ for $n=1$.
For each crossing, a jump appears in the current as a function of $\varphi$, as shown in figure~\ref{fig:Sub_7}~(a). The persistent current is therefore proportional to the level spacing at $E_{\rm F}$ for each flux value, i.e., it is small for a large density of states and large for a small density of states and vanishes in the limit of a continuous density of states (c.f. reference~\cite{loder:09}). The amplitude of the normal persistent current is thus a measure for the difference of the energy spectrum at integer and half integer flux values. It is maximal for the one dimensional case, but might be very small in real metal loops (c.f. measurements of the Aharonov-Bohm effect in metal rings~\cite{AB}).

Upon entering the superconducting state, an energy gap develops around $E_{\rm F}$ preventing energy levels from crossing $E_{\rm F}$. Thus the persistent supercurrent arises as in the one dimensional ring independent of the density of states. The large jumps in the supercurrent appear at the value of $\varphi$ where the energies of the even-$q$ and odd-$q$ states become degenerate and $q$ switches to the next integer. In the flux regimes with no crossings of $E_{\rm F}$ the supercurrent is linear and the total energy quadratic in $\varphi$. For the largest value shown ($\Delta=0.006\,t$), there is a direct gap for all values of $\varphi$. Even for this large $\Delta$, the current and the energy are not precisely $\Phi_0/2$-periodic because of the energy difference of the even and odd $q$ states in finite systems \cite{vakaryuk:08,loder:08}. The offset of the $q$-jump is only relevant for values of the pairing interaction $V$ for which $\Delta$ is finite for all $\varphi$. In figure~\ref{fig:Sub_7}~(b), the offset is clearly visible for the largest two values of $V$ (marked with black arrows). Its sign depends on the shape of the annulus and the value of $V$--- the offset changes sign for increasing $V$ (cf.\ reference~\cite{vakaryuk:08}).

\begin{figure*}[t]	
\centering
\vspace{3mm}
\begin{overpic}
[width=0.95\columnwidth]{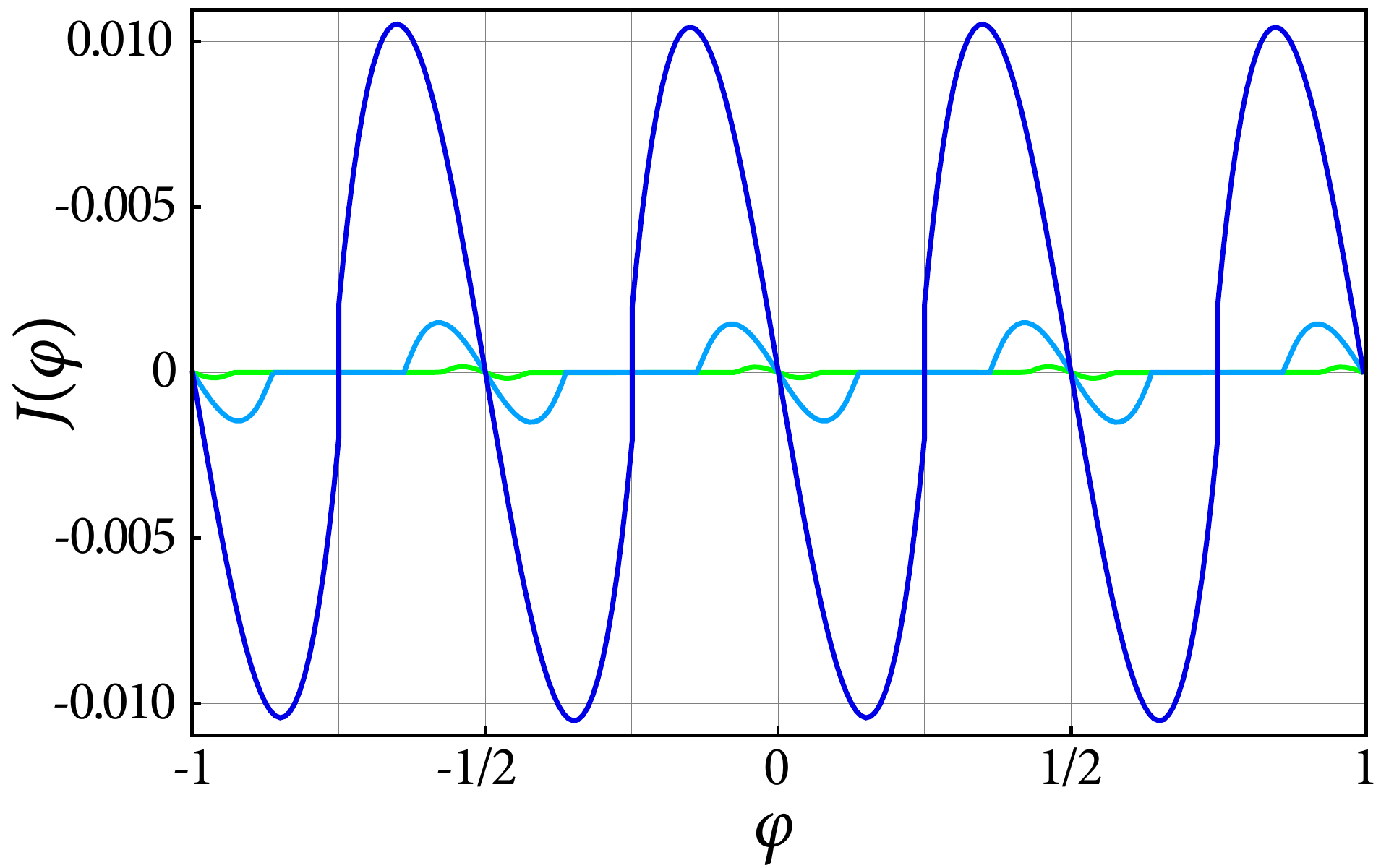}
\put(-3,59){(a)}
\end{overpic}
\hspace{6mm}
\begin{overpic}
[width=0.95\columnwidth]{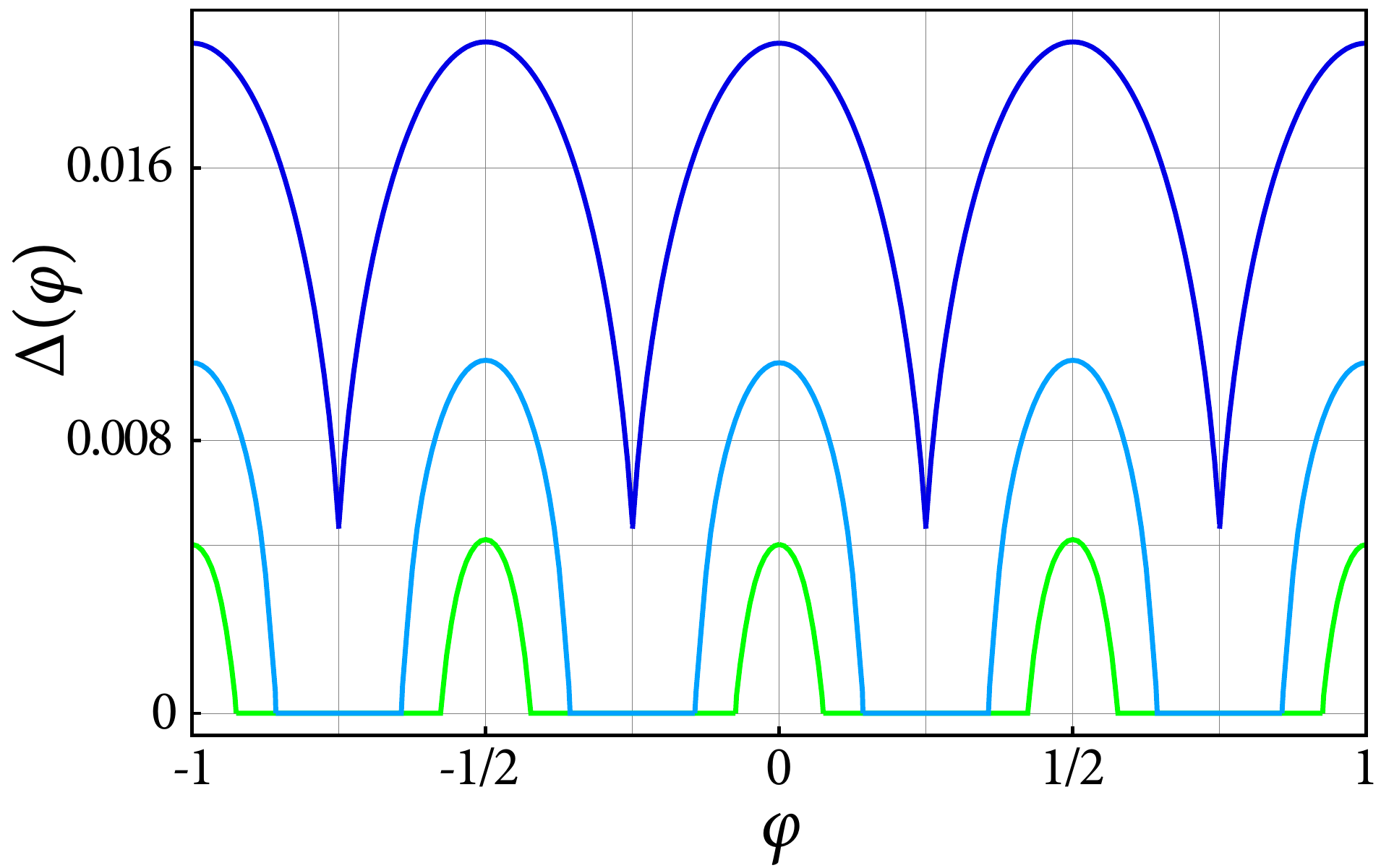}
\put(-3,59){(b)}
\end{overpic}\\[0mm]
\caption{The order parameter $\Delta(\varphi)$ and the persistent current $J(\varphi)$ for the temperature driven transition from the normal to the superconducting state in an annulus with inner radius $R_1=30a_\perp$ and outer radius $R_2=36a_\perp$. The pairing interaction is $V_0=0.7\,t$, with a critical temperature of $k_{\rm B}T_{\rm c}\approx0.0523\,t$  for zero flux. For these parameters $\Delta(T=0)\approx0.1\,t$. The lines (from top to bottom) correspond to the temperatures $k_{\rm B}T= 0.0513\,t$ (blue), $k_{\rm B}T= 0.0520\,t$ (turquoise), and $k_{\rm B}T= 0.0522\,t$ (green). Notice that $\Delta$ is slightly different for the flux values $\varphi=0$ and $\varphi=\pm1/2$.}
\label{fig:Sub_8}
\end{figure*}

The introduction of self-consistency for the order parameter does not fundamentally change these basic observations [figure~\ref{fig:Sub_7}~(b)]. One finds that $\Delta(R_1)<\Delta(R_2)$, but if $(R_2-R_1)/R_1\lesssim1$, the difference is small. In the following, we denote the average of $\Delta(r)$ by $\Delta$. The crossover is then controlled by the pairing interaction strength $V$, for which we chose such values as to reproduce the crossover from the normal state to a state with direct energy gap for all flux values. The order parameter $\Delta$ is now also a function of $\varphi$. If $\Delta(\varphi=0)\lesssim 0.006\,t$ [c.f.\ figure~\ref{fig:Sub_7}~(b)], the gap closes with $\varphi$, and $\Delta$ decreases whenever a state crosses $E_{\rm F}$. Unlike in a one dimensional ring, $\Delta$ does not drop to zero at the closing of the energy gap, but decreases stepwise. This is because in two or three dimensions, $\Delta$ is stabilized  beyond the depairing velocity by contributions to the condensation energy from pairs with relative momenta perpendicular to the direction of the current flow; the closing of the indirect energy gap does not destroy superconductivity \cite{bardeen:62, zagoskin}. 

Experimentally more relevant is to control the  crossover through temperature. With the pairing interaction $V$ sufficiently strong to produce a $T=0$ energy gap much larger than the maximum Doppler shift, the crossover regime is reached for temperatures slightly below $T_{\rm c}$. For the annulus of figure~\ref{fig:Sub_6}, the crossover proceeds within approximately one percent of $T_{\rm c}$. The crossover regime becomes narrower for larger rings, proportional to the decrease of the Doppler shift. In the limit of a quasi one-dimensional ring of radius $R$ we can be more precise: If we define the crossover temperature $\displaystyle T^*$ by $\displaystyle \Delta(T^*)=\Delta_{\rm c}$ and assuming $\Delta_{\rm c}\ll\Delta$, we can use the Ginzburg-Landau form of the order parameter
\begin{align}
\frac{\Delta(T)}{\Delta(0)}\approx1.75\sqrt{1-\frac{T}{T_{\rm c}}}
\label{41}
\end{align}
and obtain
\begin{align}
\frac{T_{\rm c}-T\displaystyle^*}{T_{\rm c}}&\approx\frac{\Delta_{\rm c}^2}{3.1\Delta^2(0)}=\frac{t^2}{12.4\Delta^2(0)(R/a)^2}\nonumber\\&=\frac{E_{\rm F}^2}{3.1(k_{\rm B}T_{\rm c})^2(R/a)^2},
\label{37}
\end{align}
where we used the relations $\Delta_{\rm c}=at/2R$ and $E_{\rm F}=2t$ for a discretized one-dimensional ring at half filling, and the BCS relation $\Delta^2(0)\approx3.1(k_{\rm B}T_{\rm c})^2$.
For a ring with a radius of 2500 lattice constants ($\approx10\,\upmu$m) and $\Delta(0)=0.01\,t$ ($\approx3\,$meV) one finds the ratio $\displaystyle(T_{\rm c}-T^*)/T_{\rm c}\approx1.3\times10^{-4}$. This is in reasonable quantitative agreement with the experimental results of Little and Parks \cite{Little,Parks}, discussed also by Tinkham in reference~\cite{Tinkham}. Their prediction was similar to equation~(\ref{37}), up to a factor in which they include a finite mean free path. But they did not included the difference introduced through even and odd $q$ states. This difference was considered in calculations of $T_{\rm c}$ by Bogachek {\it et al.\/} \cite{bogachek:75} in the single-channel limit and found to be exponentially small. A detailed study of the normal- to superconducting phase boundary was also done by Wei and Goldbart in reference~\cite{wei:07} in which they considered the $\Phi_0$ periodic contributions. 
In equation~(\ref{37}) the value of $\Delta(0)$ is in fact different for even and odd $q$.
Although quantitative predictions of $T_{\rm c}-T\displaystyle^*$ of the theory presented here might be too large as compared to the experiment, it serves as an upper limit because it describes the maximum possible persistent current. Inhomogeneities and scattering processes in real systems further reduce the difference of the energy spectra in the even- and odd-$q$ flux sectors and thereby reduce $T_{\rm c}-T\displaystyle^*$.

For temperatures close to $T_{\rm c}$, the difference of the eigenenergies of even and odd $q$ states is less important than at $T=0$. Thus the deviation from the $\Phi_0/2$ periodicity of the current and of the order parameter is smaller. Furthermore, persistent currents in the normal state are exponentially small compared to the supercurrents below $T_{\rm c}$. Their respective $\Phi_0$ periodic behavior is therefore essentially invisible for the flux values where $\Delta=0$.  In figure~\ref{fig:Sub_8}, the difference between $\Delta(\varphi=0)$ and $\Delta(\varphi=1/2)$ is still visible, but the corresponding differences in the current are too small. Only for a superconductor with very small $T_{\rm c}$, we expect the periodicity crossover to be visible.

Although we found within the framework of the BCS theory, that the crossover to a $\Phi_0/2$ periodic supercurrent takes place slightly below $T_{\rm c}$, detailed  studies by Ambegaokar and Eckern~\cite{ambe:91} and by von Oppen and Riedel~\cite{vonoppen:92} including superconducting fluctuations reveals that the crossover might actually take place above $T_{\rm c}$. This fluctuation driven crossover is broader than the BCS crossover with a similarly, exponentially suppressed $\Phi_0$ periodic normal current contribution. For a superconductor with a $T_{\rm c}$ small enough to observe a normal persistent current above $T_{\rm c}$, Eckern and Schwab suggested that the crossover regime, where both $\Phi_0$ and $\Phi_0/2$ periodic current contributions are present, should be observable at a temperature $T\approx2T_{\rm c}$~\cite{eckern:94,eckern}.

The discussion of the periodicity crossover in a multi-channel loop also gives insight into the flux periodicity of loops of unconventional superconductors with gap nodes like a $d$-wave superconductor. In nodal superconductors the density of states is finite arbitrarily close to $E_{\rm F}$. Therefore some energy levels cross $E_{\rm F}$ as a function of the flux, regardless of the size of the order parameter and consequently, the ``small gap\textquotedblright\  situation extends to arbitrarily large loops~\cite{loder:08,barash:07,tesanovic:08,loder:09}. Of course, the number of energy levels crossing $E_{\rm F}$ decreases with increasing ring size and thus also the $\Phi_0$ periodic contribution to the supercurrent. The dependence of this $\Phi_0$ periodic contribution on the ring size depends on the order parameter symmetry. The careful study in reference~\cite{loder:09} revealed that for $d$-wave superconductors, the relation between the $\Phi_0$ and $\Phi_0/2$ periodic current contributions is proportional to $1/R_1$. It was estimated that for a ring of a cuprate superconductor with a circumference of $\sim1\,\upmu$m, this ratio is about 1\% and should be observable experimentally.

\section{Conclusions}

We analyzed the crossover from the $\Phi_0$ periodic persistent currents as a function of magnetic flux in a metallic loop to the $\Phi_0/2$ periodic supercurrent in the groundstate of the loop. We considered conventional $s$-wave pairing in a one-dimensional as well as in a multi-channel annulus. 
Although a one-dimensional superconducting ring is a rather idealized system, it proves valuable for discussing the physics of this crossover, which includes the emergence of a new minimum in the free energy for odd center-of-mass angular momenta $\hslash q$ of the Cooper pairs and the restoration of the flux periodicity of the free energy. The physical concepts, which we illustrated in a simplified form in section~\ref{sec:em}, remain thereby valid even in the more complex context of the self consistent calculations on the annulus.

In the superconducting state, a distinguished minimum in the free energy develops at $\varphi=q/2$. Choosing the proper value for $q$ at each flux value leads to a series of minima at integer and half-integer flux values which, however, differ in energy for finite systems.
In rings with a radius smaller than half the superconducting coherence length, the two electrons forming a Cooper pair are not forced to circulate the ring as a pair, and the supercurrent shows a $\Phi_0$ periodicity. Only if the order parameter $\Delta$ is larger than the maximal Doppler shift $\epsilon_{\rm D}$, the supercurrent is $\Phi_0/2$ periodic. This is equivalent to the condition that the maximum flux induced current is smaller than the critical current $J_{\rm c}$. 
Assuming that the relations obtained from the one-dimensional model remain valid on a ring with 
finite thickness $R_2-R_1\ll R_1$, as indeed suggested by the multi-channel model, the critical radius to observe $\Phi_0/2$ periodicity, $R_{\rm c}=at/2\Delta$, would be of the order of $1\,\upmu$m for aluminum rings.
Within the temperature controlled crossover upon cooling through $T_{\rm c}$, $\Phi_0$ periodicity might by difficult to observe since the differences in the energy spectra for integer and half-integer flux values are exponentially suppressed by temperature.  $\Phi_0$ periodicity is therefore only observable if a normal persistent current would be observable at the same temperature if superconductivity was absent.

In the introduction we referred to experiments where flux oscillations with ``fractional periodicities\textquotedblright, i.e., fractions of $\Phi_0/2$, were observed. Among various suggested origins, there is one particularly elegant approach based on a standard two-electron interaction. Consider the order parameter $\Delta_q(
\varphi)$ for electron pairs with center-of-mass angular momentum $\hslash q$. In real space, $q$ describes the phase winding of the order parameter $\Delta(\theta,\varphi)=\Delta(\varphi) e^{iq\theta}$, where $\theta$ is the angular coordinate in the ring and $\Delta(\varphi)$ is real. To ensure that $\Delta(\theta,\varphi)$ is a single valued and continuous function, $q$ must be an integer number. If, however, $\Delta(\theta,\varphi)$ is zero somewhere on the ring, it can change sign. Such a sign changing order parameter is modeled as
\begin{multline}
\tilde\Delta(\theta,\varphi)=\frac{1}{2}\left[\Delta_0(\varphi)+\Delta_q(\varphi)e^{iq\theta}\right]\\\xrightarrow{\Delta_q=\Delta_0}\Delta(\varphi)e^{-iq\theta/2}\cos\left(\frac{q}{2}\theta\right),
\label{intr1}
\end{multline}
which displays a phase-winding number $q/2$ if $\Delta_q=\Delta_0$, and consequently a vanishing supercurrent at the fractional flux value $q\Phi_0/4$. In momentum space, $\tilde\Delta(\theta,\varphi)$ is represented by the two-component order parameter $\{\Delta_0(\varphi),\Delta_q(\varphi)\}$. Such a superconducting state is typically referred to as a ``pair-density wave\textquotedblright\  (PDW) state~\cite{agterberg:08}, since the real-space order parameter is periodically modulated and therefore $q$ can no longer be interpreted as an angular momentum. Agterberg and Tsunetsugu showed within a Ginzburg-Landau approach, that a PDW superconductor indeed allows for vortices carrying a $\Phi_0/4$ flux quantum~\cite{agterberg:08,agterberg:09}.

Based on a microscopic model, it was shown in reference~\cite{loder:10} that a PDW state cannot result from an on-site pairing interaction. However, the PDW state can be a stable alternative for unconventional superconductors with gap nodes, specifically for $d$-wave superconductivity as realized in the high-$T_{\rm c}$ cuprate superconductors. More work is, however, needed to analyze under which conditions the PDW state indeed develops an energy minimum at multiples $\Phi_0/4$, and these minima become degenerate in the limit of large loops.
The notion of the absence of coexistence of Cooper pairs with different center-of-mass momenta for conventional superconductors (as, e.g., in the PDW state) justifies the reduction of the sum over $q$ in the Hamiltonian~(\ref{H0}) in section~\ref{sec:em} to one specific $q$ in order to derive the periodicity crossover in superconducting rings.

Here we mention a further system where a similar mechanism as described above may lead to $\Phi_0/4$ flux periodicity: Sr$_2$RuO$_4$. Experimental evidence exists that  Sr$_2$RuO$_4$ is a spin triplet $p$-wave superconductor. A triplet superconductor can be represented by the two-component order parameter $\{\Delta_{q_1}^{\ua\ua}(\varphi),\Delta_{q_2}^{\da\da}(\varphi)\}$, with the center-of-mass momenta $q_1$ and $q_2$ for the $s_z=1$ and the $s_z=-1$ condensates. This realizes a similar situation as for the PDW state where $\Phi_0/4$ periodicity is possible~\cite{vakaryuk:11}. Indeed Jang \etal\ observed recently that the flux through a microscopic Sr$_2$RuO$_4$ ring is quantized in units of $\Phi_0/4$~\cite{jang:11}.

\section*{Acknowledgements}

The authors acknowledge discussions with Yuri Barash, Ulrich Eckern, Jochen Mannhart, and Christoph Schneider. This work was supported by the Deutsche Forschungsgemeinschaft through TRR~80.

\enlargethispage{10pt}
\bibliography{loder}
 
\end{document}